\documentclass[12pt,preprint]{emulateapj}

\makeatother

\newcommand{\s}{{\it Spitzer}}
\newcommand{\h}{{\it Herschel}}
\newcommand{\msol}{$\rm M_{\odot}$}
\newcommand{\md}{$M_{\rm dust}$}

\newcommand{\lsol}{L$_{\odot}$}
\newcommand{\lir}{$L_{\rm IR}$}

\newcommand{\lco}{$L^{\prime}_{\rm CO}$}

\newcommand{\td}{$T_{\rm d}$}

\newcommand{\mgas}{$M_{\rm gas}$}

\newcommand{\lcii}{$L_{\rm CII}$}
\newcommand{\lfir}{$L_{\rm FIR}$}
\newcommand{\Mmol}{$M_{\rm H_{\rm 2}}$}

\lefthead{Magdis et al..~}

\slugcomment{to appear in ApJ}
\shorttitle{The Evolving ISM of ULIRGs}
\begin{document}
 \title{A Far-Infrared Spectroscopic Survey of Intermediate Redshift (Ultra) Luminous Infrared Galaxies }
 \author{Georgios E. Magdis $\!$\altaffilmark{1,2},
 	     D. Rigopoulou $\!$\altaffilmark{1,3},
		 R. Hopwood $\!$\altaffilmark{4},
		J.-S Huang $\!$\altaffilmark{5,6,7},      
 	     D. Farrah $\!$\altaffilmark{8},
	     C. Pearson $\!$\altaffilmark{3,9},
	     Almudena Alonso-Herrero  $\!$\altaffilmark{10},
	     J. J. Bock $\!$\altaffilmark{11,12},
	     D. Clements $\!$\altaffilmark{4},
	     A. Cooray $\!$\altaffilmark{12,13},
             M. J. Griffin $\!$\altaffilmark{14},
	     S. Oliver $\!$\altaffilmark{15},
	     I. Perez Fournon $\!$\altaffilmark{16,17},
	     D. Riechers $\!$\altaffilmark{18},
	     B.M. Swinyard $\!$\altaffilmark{2,19}
	     D. Scott $\!$\altaffilmark{20},
	     N. Thatte $\!$\altaffilmark{2},
	     I. Valtchanov $\!$\altaffilmark{21},
	     M. Vaccari  $\!$\altaffilmark{22}}          
\altaffiltext{1}{Department of Physics, University of Oxford, Keble Road, Oxford OX1 3RH, UK}
\altaffiltext{2}{Institute for Astronomy, Astrophysics, Space Applications and Remote Sensing, National Observatory of Athens, GR-15236 Athens, Greece}
\altaffiltext{3}{RAL Space, Science, and Technology Facilities Council, Rutherford Appleton Laboratory, Didcot OX11 0QX, UK}
\altaffiltext{4}{Physics Department, Imperial College London, South Kensington Campus, London SW7 2AZ, UK}
\altaffiltext{5}{National Astronomical Observatories of China, Chinese Academy of Sciences, Beijing 100012, China}
\altaffiltext{6}{China-Chile Joint Center for Astronomy, Chinese Academy of Sciences, Camino El Observatorio, 1515, Las Condes, Santiago, Chile}
\altaffiltext{7}{Harvard-Smithsonian Center for Astrophysics, 60 Garden Street, Cambridge, MA 02138, USA}
\altaffiltext{8}{Department of Physics, Virginia Tech, Blacksburg, VA 24061, USA}
\altaffiltext{9}{The Open University, Milton Keynes MK7 6AA, UK), AC(Department of Physics, Virginia Tech, Blacksburg, VA 24061, USA}
\altaffiltext{10}{Instituto de Fisica de Cantabria, CSIC-UC, E-39006 Santander, Spain}
\altaffiltext{11}{California Institute of Technology, 1200 E. California Blvd., Pasadena, CA 91125, USA}
\altaffiltext{12}{Jet Propulsion Laboratory, 4800 Oak Grove Drive, Pasadena, CA 91109, USA}
\altaffiltext{13}{Department of Physics and Astronomy, University of California, Irvine, CA 92697, USA}
\altaffiltext{14}{School of Physics and Astronomy, Cardiff University, Queens Buildings, The Parade, Cardiff CF24 3AA, UK}
\altaffiltext{15}{Astronomy Centre, Department of Physics and Astronomy, University of Sussex, Brighton BN1 9QH, UK}
\altaffiltext{16}{Instituto de Astrof'sica de Canarias (IAC), 38200, La Laguna, Tenerife, Spain ipf@iac.es} 
\altaffiltext{17}{Departamento de Astrof'sica, Universidad de La Laguna (ULL), 38205, La Laguna, Tenerife, Spain}
\altaffiltext{18}{Department of Astronomy, Cornell University, 220 Space Sciences Building, Ithaca, NY 14853, USA}
\altaffiltext{19}{Dept. of Physics \& Astronomy, University College London, Gower}
\altaffiltext{20}{Department of Physics and Astronomy, University of British Columbia, 6224 Agricultural Road, Vancouver, BC V6T1Z1, Canada}
\altaffiltext{21}{Herschel Science Centre, European Space Astronomy Centre, Villanueva de la Canada, E-28691 Madrid, Spain}
\altaffiltext{22}{Astrophysics Group, Physics Department, University of the Western Cape, Private Bag X17, 7535 Bellville, Cape Town, South Africa}

\begin{abstract}
 
We present \h\ far-IR photometry and spectroscopy as well as ground based CO observations of an intermediate redshift (0.21 $\leq$  $z$ $\leq$ 0.88) sample of \h-selected (ultra)-luminous infrared galaxies (\lir\ $>$ 10$^{11.5}$\,\lsol). With these measurements we trace the dust continuum, far-IR atomic line emission, in particular [CII]\,157.7$\,\mu$m, as well as the molecular gas of $z\sim 0.3 $ (U)LIRGs and perform a detailed investigation of the interstellar medium of the population. We find that the majority of \h-selected intermediate redshift (U)LIRGs have  \lcii/\lfir\ ratios that are a factor of about 10 higher than that of local ULIRGs and comparable to that of local normal and high-$z$ star forming galaxies. Using our sample to bridge local and high$-z$ [CII] observations, we find that the majority of galaxies at all redshifts and all luminosities follow a \lcii$-$\lfir\ relation with a slope of unity, from which local ULIRGs and high$-z$ AGN dominated sources are clear outliers. We also confirm that the  strong anti-correlation between the \lcii/\lfir\ ratio and the far-IR color $L_{\rm 60}$/$L_{\rm 100}$ observed in the local Universe holds over a broad range of redshifts and luminosities, in the sense that warmer sources exhibit lower \lcii/\lfir\ at any epoch. Intermediate redshift ULIRGs are also characterised by large molecular gas reservoirs and by lower star formation efficiencies compared to that of local ULIRGs. The high \lcii/\lfir\ ratios, the moderate star formation efficiencies (\lir/\lco\ or \lir/\Mmol) and the relatively low dust temperatures of our sample (which are also common characteristics of high$-z$ star forming galaxies with ULIRG-like luminosities) indicate that the evolution of the physical properties of (U)LIRGs between the present day and $z > 1$ is already significant by $z \sim 0.3$.
\end{abstract}


\section{Introduction}
Luminous and ultraluminous infrared Galaxies (LIRGs and ULIRGs), are the most powerful class of star
forming galaxies in the Universe, with bolometric infrared luminosities (\lir) in the range of 10$^{11}$\,\lsol\ $<$ \lir\ $< 10^{12}$\,\lsol\ and \lir\ $>$ 10$^{12}$\,\lsol, respectively, which correspond to star formation rates (SFR) from tens to thousands of solar masses per year. While rare at the present day, their co-moving density increases by a factor of $\sim$ 1000 as we move to higher redshifts and they account for more than 50\% of the total star formation rate density in the Universe at $z \sim 1-2$ (e.g., Le Floc'h et al.\,2005, Magnelli et al.\,2011, Murphy et al.\,2011). Being the hosts of the bulk of star formation activity during the epoch of galaxy assembly (at $z \sim 2$), then placing (U)LIRGs (\lir $>$ 10$^{11}$ \lsol) in the current framework of galaxy evolution is a cornerstone of modern astrophysics.

Besides the strong evolution in their number density, recent studies have  pointed out some distinct differences in the properties of the present day and high$-z$ ($z$ $>$ 1) ULIRGs. Local ULIRGs (z $<$  0.2) are almost always associated with a disturbed system of galaxies undergoing mergers (e.g., Armus et al.\,1987, Melnick \& Mirabel\,1990, Clements et al.\,1996, Farrah et al.,\,2001, Bushouse et al.\,2002), are predominantly, sources with warm spectral energy distributions (dust temperature, \td\, exceeding, on average, 40\,K, e.g., Clements et al.\,2010, Symeonidis et al.\,2011), high star formation efficiencies (defined as the ratio of the total infrared luminosity over the molecular gas mass, SFE = \lir/\Mmol\ $>$ 100\,\lsol/M$_{\odot}$, (e.g., Solomon et al.\,1992, Gao \& Solomon\,2004), exhibit a general deficiency of far$-$IR atomic lines with respect to local normal galaxies (e.g., Luhman et al.\,2003, Garcia-Carpio et al.\,2011, Farrah et al.\,2013, Diaz-Santos et al.\,2013), and are characterized by compact star forming regions, confined within the central kpc of the merging systems (e.g., Solomon et al.\,1997,  Bryant \& Scoville\,1999, Soifer et al.\,2000). 

However, there is strong evidence that the aforementioned properties are not representative of the ULIRG population in the early Universe. 
For example, morphological studies have shown that high$-z$ ULIRGs appear to be a mixture of merging/interacting systems and disk galaxies (e.g., Kartaltepe et al.\,2012, Kaviraj et al.\,2013). Furthermore, studies initiated with the Spitzer Space Telescope (\s, Werner et al.\,2004) and more recently with the Herschel Space Observatory (\h, Pilbratt et al.\,2010) have shown that their infrared SEDs are more similar to those of local galaxies of lower luminosities, exhibiting stronger PAH features (e.g., Farrah et al.\,2008, Pope et al.\,2006) and colder dust temperatures (e.g., Pope et al.\,2006,\,2008, Farrah et al.\,2008, Lonsdale et al.\,2009, Symeonidis et al.\,2009\,2013, Muzzin et al.\,2010, Magdis et al.\,2010,2012b, Elbaz et al.\,2011) than local ULIRGs. Observations of the carbon monoxide molecule (CO), have also revealed an extended distribution and ordered rotation of the molecular gas (\mgas), as well as moderate star formation efficiencies that are a factor of 10 lower compared to that of local ULIRGs (e.g., Genzel et al.\,2010, Daddi et al.\,2010, Magdis et al.\,2012b). Hints of an increase in the physical scale of ULIRGs at high$-z$ have also been provided by radio and mid-IR studies, suggesting that, on average, local ULIRGs are 5 times smaller (e.g., Rujopakarn et al.\,2011, Elbaz et al.\,2011). Finally, far-IR spectroscopy targeting the [CII] 157.7$\,\mu$m atomic line of high$-z$ star formation dominated ULIRGs ($z > 1$) revealed  [CII]-to-far infrared luminosity ratios (\lcii/\lfir) similar to local star-forming galaxies, and much above the median values found for local ULIRGs, suggestive of softer radiation fields and more extended star-forming regions (e.g., Ivison et al.\,2010, Hailey-Dunsheath et al.\,2010, Stacey et al.\,2010, Valtchanov et al.\,2011).

While a complete picture has not emerged yet, it is established that there has been a strong evolution in the physical properties and possibly of the mechanism (i.e. mergers versus secular evolution) that drives the star formation activity in ULIRGs, between the present day and $z > 1$. To a great extent these observations also suggest that a simple luminosity classification is perhaps only
meaningful at a given cosmic epoch and cannot be used to draw analogs between the present day and the early Universe galaxy populations. However, ULIRGs have so far only been extensively studied either in the local universe $z < 0.2$ or at high$-z$ ($z$ $>$ 1). We are therefore building the picture at the two ends of the evolution while missing the critical ÒintermediateÓ epoch of $0.2 < z < 1.0$. To fully understand the evolution of the ULIRG phenomenon, which is responsible for the bulk of star formation in the Universe, we need to sample the evolution in the $0.2 < z < 1$ redshift range, an era when the Universe experienced a strong increase in its star formation rate density. 

\begin{table*}
{\footnotesize
\caption{Sample definition}             
\label{tab:sample}      
\centering                          
\begin{tabular}{l c c c c c}        
\hline\hline                 
Source & R.A.$^{a}$ & Decl. & $z_{\rm spec}$ & Reference$^{b}$ & Type$^{c}$ \\
& (J2000) & (J2000)&&\\
\hline
CDFS2  & 03:28:18.0 &  $-$27:43:08.0 & 0.248 & Optical(1)& AGN(1)\\
BOOTES3& 14:28:49.7 & +34:32:40.3  & 0.219 & IRS(2)& SB(2)\\
BOOTES2& 14:32:34.9&  +33:28:32.3  & 0.250 & IRS(2,8)& SB(8)\\
SWIRE4 & 10:32:37.4 & +58:08:46.0  & 0.251 & IRS(3)& SB(3)\\
ELAISS & 00:40:14.6&   $-$43:20:10.1  & 0.265& Optical(4)& SB(4)\\
CDFS1  & 03:29:04.3&   $-$28:47:52.9 & 0.289 & Optical(5) &$-$\\
BOOTES1& 14:36:31.9&   +34:38:29.1 & 0.354 & IRS(6) & AGN(6) \\
SWIRE2 & 10:51:13.4&   +57:14:26.2 & 0.362 & Optical(7) &$-$ \\
SWIRE5 & 10:35:57.9&   +58:58:46.2 &  0.366 & CO,[CII](8) &$-$\\
FLS01  & 17:20:17.1&    +59:16:37.7 & 0.417 & Optical(9), IRS(10)& Composite(10)\\
FLS02  & 17:13:31.5 &  +58:58:04.2 & 0.436 & Optical-SWIFT(8), IRS(8)& AGN(8)\\
SWIRE1 & 10:47:53.3&   +58:21:06.2& 0.887 & Optical(11) & QSO(11)\\
XMM1$^{d}$ &  02:20:16.5 &   $-$06:01:43.2 & 2.308 & (12)& SB\\
HLSW-01$^{e}$ (SWIRE6) &  10:57:50.4&   +57:30:28.3  & 2.957 & CO(13) &$-$\\
\hline
ELAISN & 16:10:27.0&   +53:12:43.0 & 0.693 & $z_{\rm phot}$&$-$\\
SWIRE7 & 11:02:05.7&   +57:57:40.6 & 0.550 & $z_{\rm phot}$&$-$\\    
XMM2&  02:19:57.3&   $-$05:23:48.9 &0.280& $z_{\rm phot}$&$-$\\
\hline           
\end{tabular}\\
Notes:\\
\noindent $^{a}$ Coordinates from MIPS 24$\,\mu$m continuum emission. $^{b}$ Origin of the spectroscopic redshift: (1)Eales et al.\,(2009); (2)Weedman et al.\,(2008);  (3)Magdis et al.\,(2013); (4)La Franca et al.\,(2004); (5)Norris et al.\,(2006); (6)Houck et al.\,(2007);  (7)Oyabu et al.\,(2005); (8)This study; (9)Papovich et al.\,(2006); (10)Dasyra et al.\,(2009); (11)Gruppioni et al. priv. com.; (12)Fu et al.\,(2013); and (13) Scott et al.\,(2011). $^{c}$ Source classification. $^{d}$ In the original sample selection XMM1 was erroneously assigned a redshift of 0.502 that corresponds to the foreground galaxy of the lensed SMGs HXMM01 (Fu et al.\,2013).  $^{e}$ In the original sample selection SWIRE6 was erroneously assigned a redshift of 0.584 that corresponds to the foreground galaxy of the lensed HLSW-01 (Scott et al.\,2011).

}
\end{table*}
To bridge the gap between local and high$-z$ ULIRGs, here we perform a detailed and in depth investigation of a sample of 17 \h-selected 
(U)LIRGs in the redshift range of $0.21 < z < 0.8$, with fully characterized infrared spectral energy distributions from \s\ and \h. Using the Spectral and Photometric Imaging REceiver-Fourier Transform Spectrometer (Griffin et al.\,2010) on board \h\ we target primarily the [CII]\,157.7$\,\mu$m emission line, as well as the [NII]\,205$\,\mu$m and [OI]\,145$\,\mu$m atomic lines. These lines, and in particular  [CII]\,157.7$\,\mu$m, are important coolants of the ISM and can serve as powerful tracer of its physical conditions. As they are rapidly becoming the benchmark for the study of the cool atomic gas in local and distant ($z > 1$) galaxies, our survey offers for the first time the opportunity to study these lines, and therefore the properties of the ISM, in intermediate redshift galaxies. We note that the current sample, first presented in  Rigopoulou et al.\,(2014), is the only existing sample of (U)LIRGs in this redshift range with [CII] measurements. Our analysis is also complemented by follow-up ground based CO observations that trace the molecular gas of the sources, resulting in a unique sample of ULIRGs with far-IR spectroscopic and multi-wavelength photometric data, ideal to explore the evolution of ULIRGs.

This paper is organized as follows. In Sections 2 and 3 we present the selection criteria of the sample and the subsequent observations and data analysis, respectively. In Section 4 we derive the infrared properties of the sample, based on SED fitting, and in Section 5 we explore various correlations between the strength of the [CII] emission relative to the far-IR infrared luminosity and other parameters, such the total far infrared luminosity, the dust temperature, and the star-formation efficiency of galaxies across the cosmic time. In Section 6 we discuss the implications of our findings on the evolution of the (U)LIRGs phenomenon, while a summary of our results is given in Section 7. Throughout this article we adopt the $WMAP-$7 cosmology ($\Omega_{\rm m} = 0.273$, $\Omega_{\rm \Lambda} + \Omega_{\rm m} = 1$, $\Omega_{\rm k} = 0$ and $H_{0} = 70.4$\,km\,s$^{-1}$\, Mpc$^{-1}$; Larson et al.\,2011).

\section{The sample}
We selected targets to investigate the properties of the interstellar medium (ISM) of intermediate redshift (U)LIRGs through the detection of far-IR atomic lines, especially [CII], [NII] and [OI]. The two main criteria were: first a lower limit on the brightness in the far-IR to ensure the detection of the source against the background emission of the telescope ($\sim$ 80\,K) and the uncertainties in the instrument and telescope model, (see SPIRE-FTS handbook) and second the redshift range to ensure that the redshifted [CII] emission falls within the spectral bandwidth of FTS (i.e., $z >$ 0.21). The starting point for our target selection was the \h-SPIRE photometric catalogues from the \h\ Multi-tiered Extragalactic Survey (Oliver et al.\,2012). The HerMES survey mapped at 250$\,\mu$m 350$\,\mu$m and 500$\,\mu$m some of the most well studied fields in the sky, including, Bootes, EGS, COSMOS and HDFN. Photometric catalogs were produced using both a blind and a 24$\,\mu$m prior based source extraction technique (Roseboom et al.\,2010). Using the prior based catalogues we searched for sources with $S_{\rm 250} > 150\,$mJy in all fields covered by HERMES, producing a master 250$\,\mu$m target list (see Section 3.1). As stated above the flux cut was imposed by the telescope specifications as well as our desire to study far-IR luminous sources (\lir\ $> 10^{11}$ \lsol)\footnote{Using various templates (M82, Arp220, Chary \& Elbaz 2001)  we estimate a total infrared luminosity of log (\lir/\lsol) $>$11.0  for sources with  $S_{\rm 250} = 150\,$mJy in the redshift range of at $z=0.21 - 0.8$.}.  

Among the 250$\,\mu$m-selected targets 14 had spectroscopic redshifts with 0.21 $<$ $z_{\rm spec}$ $<$ 0.9, a redshift range where [CII]\,157.7\,$\mu$m and [NII]\,205\,$\mu$m emission is covered by FTS. The initial spectroscopic sample was further supplemented with the three brightest sources at 250$\,\mu$m and $z_{\rm phot} > 0.3$ resulting in a final ``\textit{[CII]-sample}'' of 17 galaxies. No other criteria (e.g., optical color, stellar mass, existence of an AGN) were imposed. The final sample consists of 15 sources in the redshift range of 0.219 $\leq$ $z$ $\leq$ 0.887, and two sources, XMM1, at $z_{\rm spec}$ = 2.308, and SWIRE6 at $z_{\rm spec}$ = 2.957. The latter were initially selected as the $z_{\rm spec}$ = 0.502 and $z_{\rm spec}$ = 0.584 foreground galaxies of the lensed SMGs HXMM01 (Fu et al.\,2013) and HLSW-01 (Scott et al.\,2011, Conley et al.\,2011, Gavazzi et al.\,2011, Riechers et al.\,2011) respectively. The FTS spectra of these two lensed systems will be presented in detail in two companion papers. For the rest of the sources, the spectroscopic redshifts are drawn from the literature  and are based on either optical or mid-IR spectra (see Table 1). Where available, optical line ratios (e.g. [NII]/H$\rm\alpha$), the equivalent width (EW) of the 6.2$\,\mu$m polycyclic aromatic hydrocarbon (PAH) feature (e.g., Spoon et l. 2007) or the mid-IR color based on IRAC photometry (e.g., Donley et al.\,2012)  were also used to identify four AGN among the galaxies in our sample (BOOTES1, SWIRE1,  CDFS2,  and FLS02). In total we have 17 sources, of which  14 have spectroscopic and three have photometric redshifts. The median redshift of the sample, excluding the two high$-z$ lensed systems, is  0.351.  The full sample is presented in Table 1.

\begin{table*}
{\footnotesize
\caption{Far-IR Photometry}             
\label{tab:far}      
\centering                         
\begin{tabular}{l c c c c c c}        
\hline\hline                 
Source&$S_{\rm 24}$&$S_{\rm 70}$&$S_{\rm 160}$&$S_{\rm 250}$ $^{a}$&$S_{\rm 350}$&$S_{\rm 500}$ \\
&[mJy]&[mJy]&[mJy]&[mJy]&[mJy]&[mJy]\\
\hline
BOOTES1&52.2   $\pm$  0.2 &614.0   $\pm$   1.5& 605.5 $\pm$     3.4& 367.4  $\pm$     2.1&  161.1  $\pm$    2.3& 58.5 $\pm$      3.4\\
BOOTES2&4.3   $\pm$  0.2 & 347.0   $\pm$  1.0& 294.5 $\pm$      2.5& 175.9 $\pm$      1.9& 69.3  $\pm$     2.1& 25.1  $\pm$     2.2\\
BOOTES3&8.7 $\pm$    0.4 & 244.0  $\pm$    0.8& 518.8  $\pm$     3.7& 351.4 $\pm$     1.9& 164.1$\pm$       1.9& 57.2 $\pm$       2.3 \\
CDFS1 & 2.9  $\pm$   0.1 & 93.0 $\pm$   1.6& 355.9  $\pm$    5.9& 195.4   $\pm$    2.5& 85.7  $\pm$     3.6& 41.3  $\pm$     12.5  \\
CDFS2 &5.0  $\pm$   0.1  & 199.2    $\pm$   2.2& 371.8  $\pm$     6.7& 227.6 $\pm$      2.5& 100.9 $\pm$     2.1&  33.1  $\pm$     2.5\\
ELAISS  & 3.9  $\pm$   0.4 & 56.4     $\pm$  2.1& 201.3   $\pm$    10.6& 158.9 $\pm$      2.3& 78.3   $\pm$    2.7& 28.3   $\pm$    3.0 \\
FLS01 & 3.3 $\pm$   0.4 & 65.0   $\pm$    0.3 & $-$&                                          80.6  $\pm$     2.4& 29.6  $\pm$    6.0&  $-$    \\
FLS02      & 6.2   $\pm$  0.7 & 230.1  $\pm$  0.3& 314.7  $\pm$     6.1& 172.2 $\pm$     2.4& 90.5  $\pm$     2.4& 38.8   $\pm$    3.4 \\
SWIRE1   &5.2   $\pm$  0.2 & 78.0   $\pm$   1.4& 279.2  $\pm$     6.2& 199.8  $\pm$     4.0& 139.9   $\pm$    4.8& 45.1  $\pm$     13.6  \\
SWIRE2   &2.7  $\pm$   0.04 & 42.7    $\pm$   1.1& 241.74  $\pm$     4.9& 201.5  $\pm$     4.2& 148.3 $\pm$      4.4& 69.1 $\pm$     5.5 \\
SWIRE4   & 6.1 $\pm$    0.5 & 139.2  $\pm$    1.8& 547.0   $\pm$    6.7& 173.1  $\pm$     7.3& 54.8   $\pm$    6.4& 14.6  $\pm$    7.9 \\
SWIRE5   & 5.9 $\pm$     0.4 & 96.0   $\pm$   1.8& 303.2   $\pm$    8.1& 174.1  $\pm$     4.0& 83.4  $\pm$    4.1& 45.7 $\pm$      6.1\\
HLSW-01 (SWIRE6)  &5.5   $\pm$  0.4 & 22.2  $\pm$    3.0& $-$& 425.0 $\pm$      10.0& 340.0 $\pm$      10.0& 233.0   $\pm$    10.0 \\
XMM1 $^{b}$      & 2.2  $\pm$   0.10 & 7.9 $\pm$    3.8& 102.1   $\pm$  6.0& 180.3  $\pm$    14.3 &192.1    $\pm$  15.5& 131.6  $\pm$    11.3\\
\hline
ELAISN   & 3.8  $\pm$   0.4& $-$& 245.9  $\pm$    7.2& 149.2 $\pm$      1.3& 83.6   $\pm$    1.2& 23.7 $\pm$     1.4\\ 
SWIRE7   &3.3  $\pm$   0.4&  72.4   $\pm$  1.3 & 224.5$\pm$     4.9& 191.2  $\pm$     4.9& 98.3    $\pm$   4.1 &41.7  $\pm$    4.9 \\
XMM2     &3.8  $\pm$ 0.2 & 336.0  $\pm$ 6.0 & 481.2 $\pm$ 11.2 & 188.7 $\pm$ 2.35 & 86.3 $\pm$ 5.5 & $<$15 \\
\hline           
\end{tabular}\\
Notes:\\
$^{a}$ The quoted uncertainties in the SPIRE photometry here do not take into account  the confusion noise.\\ 
$^{b}$ Extensive photometry for XMM1 can be found in Fu et al.\,(2013). Here the quoted flux density at 160$\,\mu$m corresponds to the PACS measurement, in contrast to the rest of  the sources, which corresponds to flux densities as measured by MIPS. \\ 
}
\end{table*}
\section{Observations and Data Analysis }

\subsection{Herschel and ancillary multi-wavelength photometry}
We used \h-SPIRE observations of the fields ELAIS-N1, ELAIS-S1, Lockman Hole, XMM, Bootes, CDFS, and FLS, obtained as 
part of the Herschel Multi-Tiered Extragalactic Survey (HerMES; Oliver et al.\, 2012). For our sources, we employed the photometric 
catalogs at 250, 350, and 500$\,\mu$m that were produced for each field by using a prior source extraction, guided by the position of known 
24$\,\mu$m sources. An extensive description of the cross-identification prior source extraction (XID) method is given in Roseboom et al.\,(2010, 
2012). The main advantage of this method is that reliable fluxes can be extracted close to the formal $\approx$ 4$-$5\,mJy SPIRE confusion noise 
(Nguyen et al.\,2010) by estimating the flux contributions from nearby sources within one beam. The 24$\,\mu$m prior positional information 
reduces the impact of confusion noise and so the approximate 3$\sigma$ limit for the SPIRE catalogue at 250$\,\mu$m becomes $\approx$ 9$-$15 mJy. The 
drawback of this technique is that the resulting catalogs could be missing sources without a 24$\,\mu$m counterpart, that is, 24$\,\mu$m drop-outs (see Magdis et al.\,2011). All sources were originally selected to have $S_{\rm 250} > 150$\,mJy, although a subsequent deblending analysis of FLS01 yielded a flux density of  $S_{\rm 250} = 80.6$\,mJy. 

Our targets also benefit from ancillary \s\ MIPS (24, 70, and 160$\,\mu$m) and IRAC (3.6, 4.5, 5.8, and 8.0$\,\mu$m) observations. Finally, six of our sources have been followed up with IRS spectroscopy. The IRS reduced spectra were downloaded from CASSIS.v6\footnote{http://cassis.sirtf.com/atlas/} (Lebouteiller et al.\, 2011) and PAH luminosities and equivalent widths (EW) were measured using PAHFIT (Smith et al.\,2007). The MIPS and SPIRE photometry of or our sample is summarized in Table \ref{tab:far}.

\subsection{FTS observations and data analysis }
The (U)LIRGs in our sample were observed with the SPIRE FTS
on board \h, between March 2012 and January 2013. 
The FTS observed 100 repetitions (13320 seconds total integration
time) on each target with sparse spatial 
sampling, in high spectral resolution (0.048\,cm$^{-1}$) mode. 
The SPIRE FTS measures the Fourier transform 
of the spectrum of a source using two bolometer detector arrays, simultaneously 
covering wavelength bands of 194--313\,$\mu$m (SSW) and 303--671\,$\mu$m (SLW). 

The data were reduced with the user pipeline (Fulton et al.\,2014 in prep) 
in HIPE (Ott et al.\,2010) version 11, or version 12.1 if the  
slightly wider frequency bands proved advantageous for fitting lines 
near the frequency band edges. All ULIRGs in our sample are extremely 
faint targets for the FTS, with continua $<<$ 1Jy, and therefore require processing 
beyond a standard reduction. 

For those observations taken within the first eight hours of an FTS pair of observing days, a correction was made to the telescope model subtracted within the pipeline. This adjustment is necessary due to low detector temperatures during the SPIRE cooler recycle, which is not fully accounted for by the non linearity correction for certain detectors, and therefore the spectra fail to be properly calibrated. The empirical correction applied is based on the linear correlation of detector temperature and SLW flux density of point source calibrated spectra of dark sky observations (see Swinyard et al.\, 2014 for details). For correctly calibrated data, this relationship is flat. 

A bespoke Relative Spectral Response Function (RSRF), constructed from selected long dark sky observations, was applied if found to improve the noise in the point source
calibrated product. However, for most observations, the standard RSRF was used,  which was constructed using the scans of all dark sky observations. Once reduced to the point source calibrated stage, the main issue affecting FTS observations of faint compact sources is imperfect subtraction of the telescope RSRF and, for the low frequency end of SLW, the instrument RSRF. To correct the distortions introduced by the point source calibration of this extended telescope residual, a background subtraction was performed. For FTS observations of point sources the target is positioned in the centre of the SSW detector, which overlaps with the central SLW detector. Both detector arrays are set out in a closely packed hexagonal pattern, which essentially provides concentric rings of off-axis detectors around the respective centre detector (see Swinyard et al.\,2014 for the precise layout). The first ring of SLW detectors, and the first and second rings for SSW, were smoothed to remove small scale noise and inspected to reject any outliers due to clipping, before averaging the selected spectra. This average gives a measure of the residual background, which was subtracted from the respective on-axis detectors.
\begin{figure}
\centering
\includegraphics[scale=0.285]{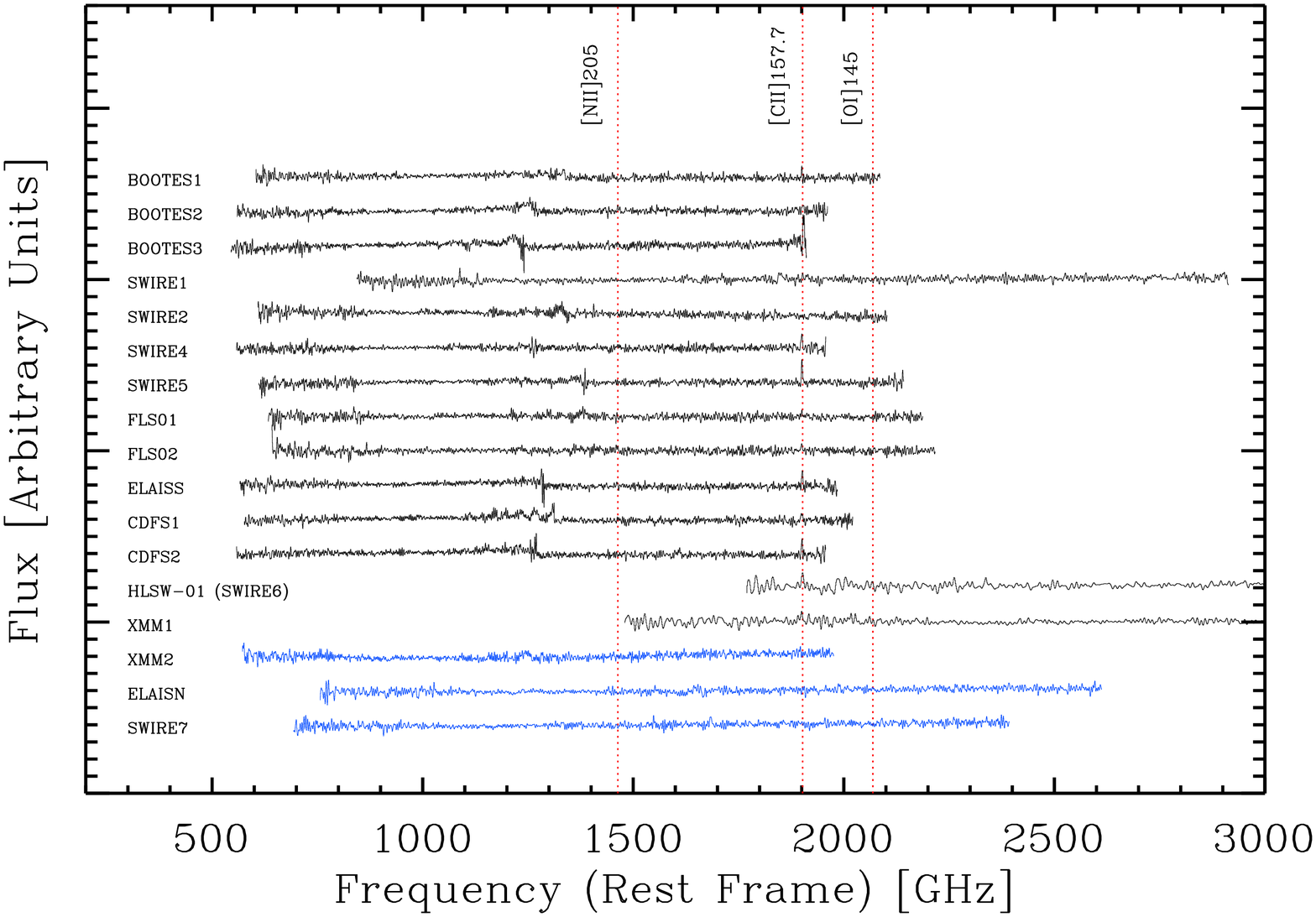}
\caption{Spectroscopic \h-FTS atlas of the sources in our sample showing the full rest-frame spectra in arbitrary units. The vertical lines correspond to the frequencies of the [NII]\,205$\,\mu$m, [CII]\,157.7$\,\mu$m and [OI]\,145$\,\mu$m atomic lines. The blue spectra correspond to sources with photometric redshift.}
\label{fig:atlas} %
\end{figure}

The optimally reduction spectra were examined for spectral features at the positions of expected lines. Due to the faint nature of the targets, many of the lines are $<\,5\,\sigma$, and since low frequency noise in an FTS spectrum can easily mimic a faint
line of similar significance, two checks were performed for reliability and 
to gain statistically robust line measurements, both utilising the
un-averaged point source calibrated data. These two methods are 
described in Hopwood et al.\,(2013) and Hopwood et al.\,(2014) in prep. 
Firstly, a Jackknife technique was used to minimise the chance 
of spurious detections. Each un-averaged set of 200 scans were divided into groups of 
sequential subsets. An average was taken for each subset, and this repeated for
subsets of decreasing number, i.e. two sets of 100 scans, 
then four sets of 50 scans, down to spectra averaged from ten scan subsets. 
In order to assess the realness of a potential detection, 
the results were plotted for a visual comparison of each averaged 
spectra in a subset, and across subset size.
Spectra of ten scans should present as random noise, unless there 
is high systematic noise, which can manifest as a detection in the fully 
averaged spectrum. The comparison of spectra averaged from twenty 
scans or more, should start to show a consistent peak at the line position for all subsets, 
while the nearby data should still present as more random, with peaks of less significance. 

Secondly, a bootstrap method was used to measure the line flux for 
any line assessed as potentially real, and to provide a background 
level, which adds a second reliability check. 
For any given observation, the un-averaged scan set was randomly 
sampled until the number in the parent population (of 200) was reached. 
These random scans were then averaged and line measurements taken 
using the same basic fitting technique as for a standard average spectrum.
The instrumental line shape of the FTS is well approximated by a sinc function 
(see Swinyard et al.\,2014 and Hopwood et al.\,2014 for more details).
Therefore, for unresolved lines, sinc functions were fitted 
simultaneously with a polynomial of order three for the continuum. 
Several of the detected [CII] lines were found to be partially resolved 
and so a sinc convolved with a Gaussian was fitted in these cases, 
using the \textsc{sincGauss} function within HIPE. 
When bootstrapping the line measurements, a more aggressive baseline 
subtraction was performed to each scan, prior to resampling. 
A set of random frequency positions were also generated and the same line fitting process applied. The resampling and fitting was repeated 10,000 
times for each observation and a Gaussian fitted to the resulting line flux 
distribution to obtain the mean line flux. The Gaussian width was taken 
as the associated 1$\sigma$ uncertainties. 
The distributions obtained from the set of random frequency positions establishes 
a background level, above which a real spectral feature is strongly suggested. 
This level was generally $\sim\,2\,\sigma$, so for any line found to be at 
$<2\,\sigma$, the bootstrapped flux density is taken as an upper limit, 
but only if the detection is supported by the Jackknife and other visual checks. 
\begin{figure}
\centering
\includegraphics[scale=0.35]{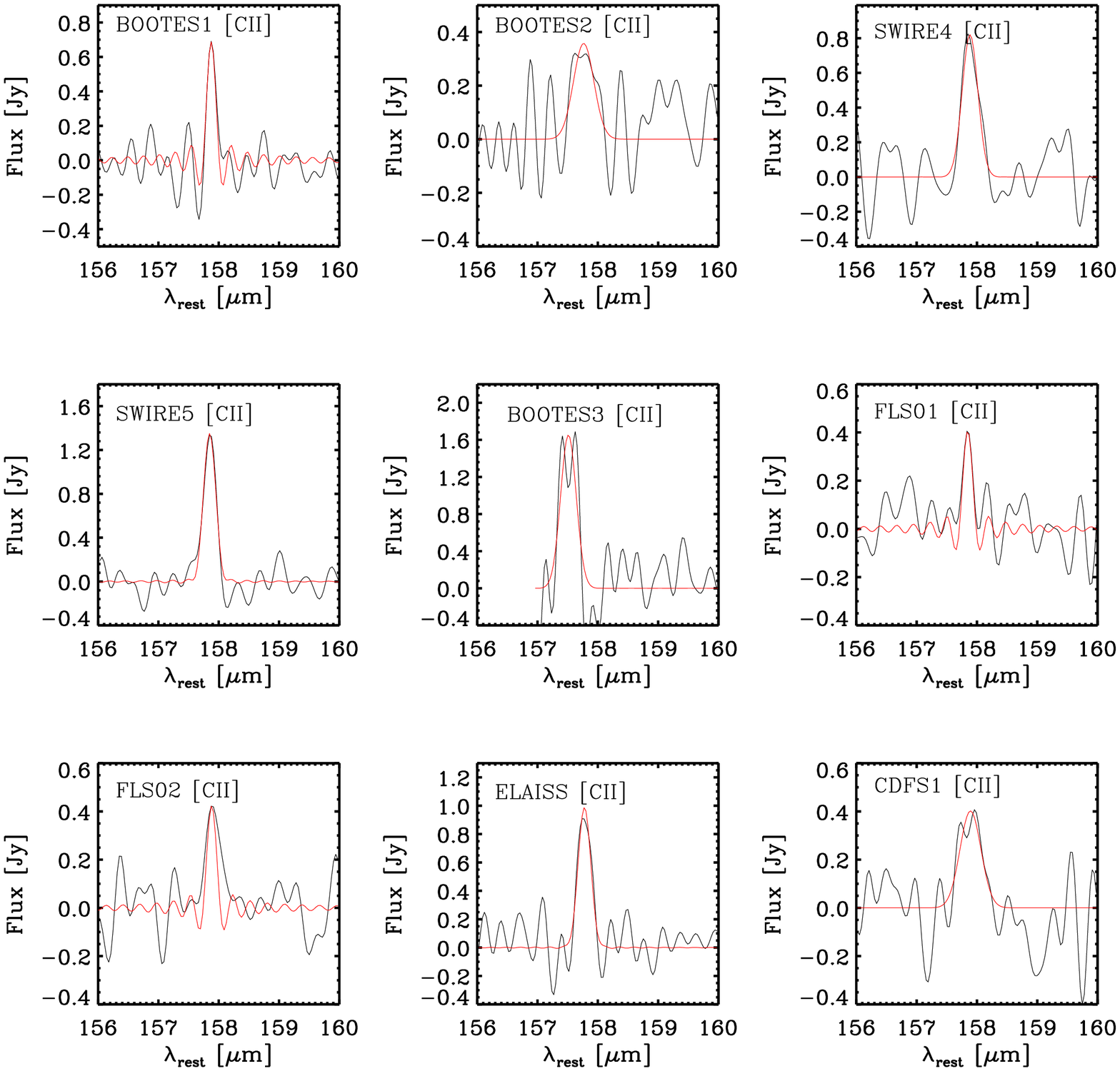}\\
\includegraphics[scale=0.36]{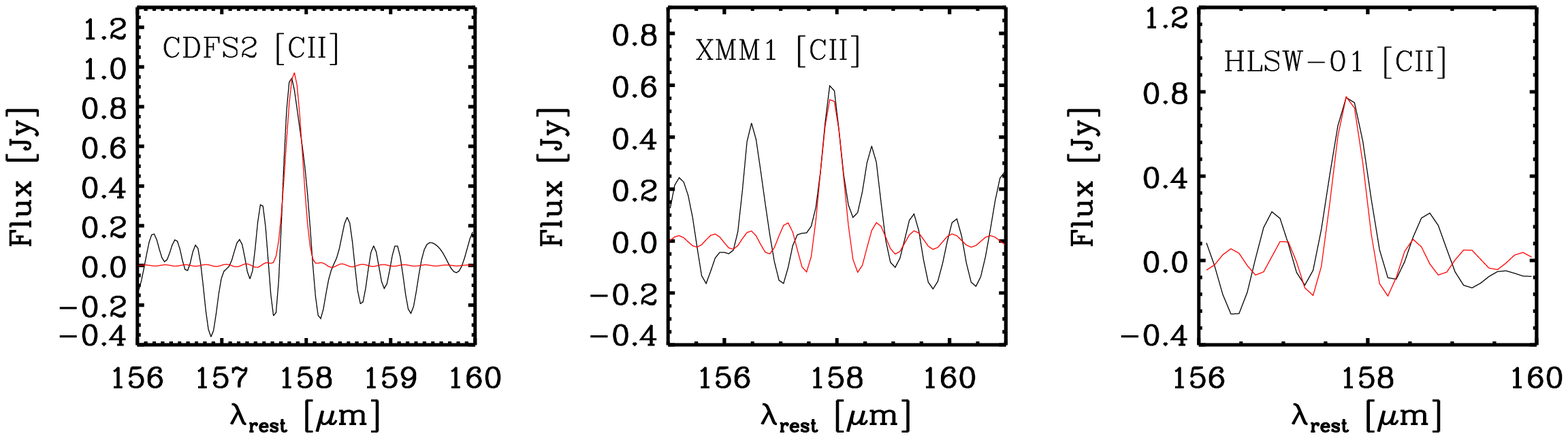}
\caption{Continuum-subtracted rest-frame spectra around the expected wavelength of the detected  [CII]\,157.7$\,\mu$m  atomic line for each source in the sample. The red curves correspond to the best fit model, which was used to extract the emission line fluxes and widths (see section 3.2).}
\label{fig:fts1} %
\end{figure}
The full extracted spectra are presented in Figure \ref{fig:atlas} while in Figure \ref{fig:fts1} we show the detected lines along with the best fit model. The derived flux densities, line widths, luminosities and 3\,$\sigma$ upper limits for the undetected lines are summarised in Table \ref{tab:atlas}. In total we detect [CII]\,157.7 emission from 12 out of the 14 sources in our spectroscopic sample. The [OI]\,145 line is also detected in one of our sources (SWIRE5) while [NII]\,205, remains undetected for the whole sample. For the three sources with photometric redshifts we do not detect any lines at the expected frequencies or by letting the redshift vary as a free parameter in the fitting process. In Table \ref{tab:atlas} we provide the upper limits of the lines measured at the expected frequencies based on their photometric redshifts. However, since the actual redshift of these sources remains largely unconstrained, we choose to  omit them from the subsequent analysis.

\begin{table*}
{\tiny
\caption{FTS Spectroscopy }             
\label{tab:atlas}      
\centering                         
\begin{tabular}{l c c c c c c c }        
\hline\hline                 
Source& $F_{\rm [OI]\,145}$$^{a}$& log $L_{\rm [OI]\,145}$$^{b}$ &$F_{\rm [CII]}$&FWHM$^{c}$& log $L_{\rm [CII]}$ &$F_{\rm [NII]\,205}$ &log $L_{\rm [NII]\,205}$  \\
&[$10^{-18}$ W m $^{-2}$] &  [L$_{\odot}$]&[$10^{-18}$ W m $^{-2}$] & [km s$^{-1}$]& [L$_{\odot}$]&[$10^{-18}$ W m $^{-2}$]&[L$_{\odot}$]\\
BOOTES1 & $<$ 6.6 & $<$ 8.85 & 7.6 $\pm$ 2.4 & 305 & 8.91$\pm$  0.13  &  $<$ 7.2 &  $<$ 8.89\\
BOOTES2 &   $-$           &  $-$ & 12.5 $\pm$ 3.5& 661 $\pm$ 149 &    8.61 $\pm$ 0.52 &     $<$  4.8 &     $<$  8.42\\
BOOTES3 &   $-$           &  $-$ &   47.7 $\pm$ 5.6 &   545 $\pm$ 43 &   9.22 $\pm$ 0.04  &   $<$  6.0&      $<$  8.32\\
CDFS1      &  $-$ &  $-$ &       16.7 $\pm$ 2.9 &764 $\pm$ 132 &      9.41 $\pm$ 0.06 &  $<$  4.8 &  $<$  8.57\\ 
CDFS2      &  $-$ &  $-$ &      20.5  $\pm$  2.8& 390 $\pm$ 54&      9.04 $\pm$  0.05 &  $<$  6.0&  $<$  8.46\\
ELAISS   &  $-$&  $-$ &       22.1  $\pm$ 2.1 &  433 $\pm$ 41  &   9.13   $\pm$  0.03   & $<$ 4.8& $<$ 8.43\\
FLS01     & $<$ 7.8 &$<$ 9.1 &  3.8 $\pm$  1.10 & 317 &  8.78 $\pm$  0.12 & $<$ 5.4& $<$  8.94\\
FLS02      & $<$ 5.7 &$<$ 9.0 &      4.7  $\pm$   1.10& 317 &     8.92 $\pm$   0.10& $<$ 4.8& $<$  8.93\\
SWIRE1   & $<$ 0.15 &$<$ 8.1 &    $<$  7.2 & $- $&   $<$  9.86 & $<$ 0.06 & $<$   7.78\\
SWIRE2   & $<$ 6.3 &$<$ 8.6 &    $<$  9.3 &  $-$&  $<$   9.03 & $<$ 5.1 & $<$  8.77\\
SWIRE4   &  $-$&  $-$ &   24.2 $\pm$  3.5 &556 $\pm$ 75  &     9.04 $\pm$  0.04 & $<$ 1.5 & $<$ 7.86\\
SWIRE5   & 4.04 $\pm $0.12& 8.6  $\pm$   0.01 &  27.8  $\pm$  2.1 &417 $\pm$ 43 &  9.41 $\pm$    0.03  & $<$ 4.5 & $<$      8.72\\
HLSW-01 (SWIRE6)   & $<$ 10.8 $\mu$  &$<$ 11.3&  (8.5  $\pm$  1.5)/$\mu$ &896 &  11.21 $\pm$    0.07  &$-$ & $-$\\
XMM1       & $<$ 11.1&$<$ 11.0 &  5.4 $\pm$  1.8& 740&       10.76 $\pm$  0.14&$-$&$-$\\
\hline
ELAISN$^{d}$     & $-$ &$-$& $<$ 5.40 &$-$&       $<$ 9.47&$-$&$-$\\
SWIRE7$^{d}$   & $<$ 11.1 &$<$   9.54 &$<$ 5.40&$-$ &$<$      9.22  & $<$ 5.4&$<$      9.22\\
XMM2$^{d}$     & 3.2 &    8.78&$<$  3.9 &$-$&$<$8.39  & $<$ 6.0 & $<$ 8.58\\
\hline
\hline           
\end{tabular}\\
Notes:\\
$^{a}$ In the cases of non-detections we report 3$\sigma$ upper limits. No value, indicates that the line is out of the FTS bandwidth for the redshift of the source.\\
$^{b}$ In the cases of non detection we report 3$\sigma$ upper limits.\\
$^{c}$ Uncertainties are quoted for sources where a gauss+sinc profile provides the best fit to the data. For the rest of the sources we quote the FWHM that corresponds to the fixed FWHM $=$ 1.4305\,GHz of the sinc function. \\
$^{d}$ Sources with photometric redshift. The upper limits correspond to measurements at the expected frequency of the lines based on the photo$-z$ estimate.
}

\end{table*}

\subsection{CO Observations and data analysis}
We carried out follow-up, CO single dish observations for nine galaxies in our sample. We  used the Eight MIxer Receiver (EMIR, Carter et al.\,2012) on the IRAM 30m telescope at Pico Veleta, Spain, in June 2012 for six northern sources  and the Swedish Heterodyne Facility Instrument (SHFI; Vassilev et al.\,2008) on the Atacama Pathfinder EXperiment {\footnote{This publication is based on data acquired with the Atacama Pathfinder Experiment (APEX). APEX is a collaboration between the Max-Planck-Institut fur Radioastronomie, the European Southern Observatory, and the Onsala Space Observatory}} (APEX, Gusten et al.\,2006) 12 m telescope, from August to October 2012 (run ID 090.B-0708A) and from April to July 2013 (run ID 091.B-0312A )  for three southern sources (CDFS1, CDFS2, and  ELAISS).
 \begin{table*}
{\footnotesize
\caption{CO IRAM and APEX Observations}             
\label{tab:co}      
\centering                         
\begin{tabular}{l c c c c c c c}        
\hline\hline                 
Source& $z_{\rm CO}$$^{a}$& Line & $S_{\rm CO} \Delta v$ & $\Delta V_{\rm FWHM}$ & log $L^{\prime}_{\rm CO}$ &log$L_{\rm IR}$ & log $L^{\prime}_{\rm CO[1-0]}$$^{b}$ \\
& & &[J km s$^{-1}$] & [km s$^{-1}$] & [K km $s^{-1}$ pc$^{2}$] & [L$_{\odot}$]& [K km $s^{-1}$ pc$^{2}$]\\
\hline
BOOTES1&0.352  & CO[1$-$0] & 1.63 $\pm$ 0.28 & 184 $\pm$ 43  &10.21 $\pm$ 0.08 & 12.69&10.21 $\pm$ 0.08\\
&0.352  & CO[2$-$1] & 7.58 $\pm$ 1.10 & 416 $\pm$ 71  & 10.09 $\pm$ 0.06 &$_{''}$&$_{''}$\\
BOOTES2&0.249 & CO[1$-$0] & 2.01 $\pm$ 0.38 & 378 $\pm$ 74  & 9.81 $\pm$ 0.08 &   11.80&9.81 $\pm$ 0.08\\
BOOTES3&0.216 & CO[1$-$0] & 6.19 $\pm$ 0.81 & 627 $\pm$ 82  & 10.17 $\pm$ 0.05 & 11.87&10.17 $\pm$ 0.05\\
SWIRE4&0.248  & CO[1$-$0] & 3.40 $\pm$ 0.57 & 556 $\pm$ 105 & 10.04 $\pm$ 0.07 & 11.79&10.04 $\pm$ 0.07\\
SWIRE5&0.366  & CO[1$-$0] & 2.78 $\pm$ 0.52 & 476 $\pm$ 145 & 10.29 $\pm$ 0.08 & 12.06&10.29 $\pm$ 0.08\\
      &0.367  & CO[2$-$1] & 8.20 $\pm$ 1.15 & 531 $\pm$ 104 & 10.16 $\pm$ 0.06 &$_{''}$&$_{''}$\\
FLS02 &0.436   & CO[2$-$1] & 3.36 $\pm$ 0.81 & 347 $\pm$ 124 & 9.93 $\pm$  0.04 & 12.41& 10.05 $\pm$ 0.05\\
      &0.436   & CO[3$-$2] & 5.76 $\pm$ 1.10 & 380 $\pm$ 77  & 9.81 $\pm$  0.08 &$_{''}$&10.11 $\pm$  0.16\\
CDFS1 &0.289  & CO[3$-$2] & 11.15 $\pm$ 2.29 & 290 $\pm$ 64 & 9.72 $\pm$ 0.08 & 11.79 & 10.03 $\pm$ 0.16\\
CDFS2 &0.248   & CO[3$-$2] & 17.05 $\pm$ 5.57 & 423 $\pm$ 198 & 9.77 $\pm$ 0.14 & 11.82& 10.07 $\pm$ 0.28\\
ELAISS&0.265   & CO[3$-$2] & 12.13 $\pm$ 3.60 & 273 $\pm$ 150 & 9.68 $\pm$ 0.12 & 11.59& 9.98 $\pm$ 0.24\\ 
\hline           
\end{tabular}\\
Notes:\\
$^{a}$ Typical uncertainty, $\Delta z = 0.001$\\
$^{b}$ For sources where CO[1-0] observations are not available we adopt $r_{\rm 21}$ = 0.75 and $r_{\rm 32}$ = 0.5.
}
\end{table*}

IRAM observations where carried out at 1.2- 2.0- and 3.0\,mm (E0, E1 and E2 bands of EMIR) targeting primarily the CO[1$-$0] emission ($\nu_{\rm 0} = 115.271$\,GHz) line from our galaxies. If  CO[1$-$0] was not accessible we searched for CO[2$-$1] ($\nu_{\rm 0} = 230.538$\,GHz) or CO[3$-$2] ($\nu_{\rm 0} = 345.796$\,GHz). When possible, we carried out simultaneous observations of two lines using E0/E1 (CO[1$-$0], CO[2$-$1]) or E1/E2 (CO[2$-$1], CO[3$-$2]) EMIR band combinations. The broadband EMIR receivers were tuned in single sideband mode, with a total bandwidth of 4\,GHz per polarization which covers a velocity range of around 12000\,km\,s$^{-1}$1 at 3\,mm,  8000\,km\,s$^{-1}$ at 2\,mm and $\sim$ 5000\,km\,s$^{-1}$ at 1.2\,mm. The observations were carried out in wobbler switching mode, with reference positions offset by 2$'$ in azimuth.

\begin{figure}
\centering
\includegraphics[scale=0.36]{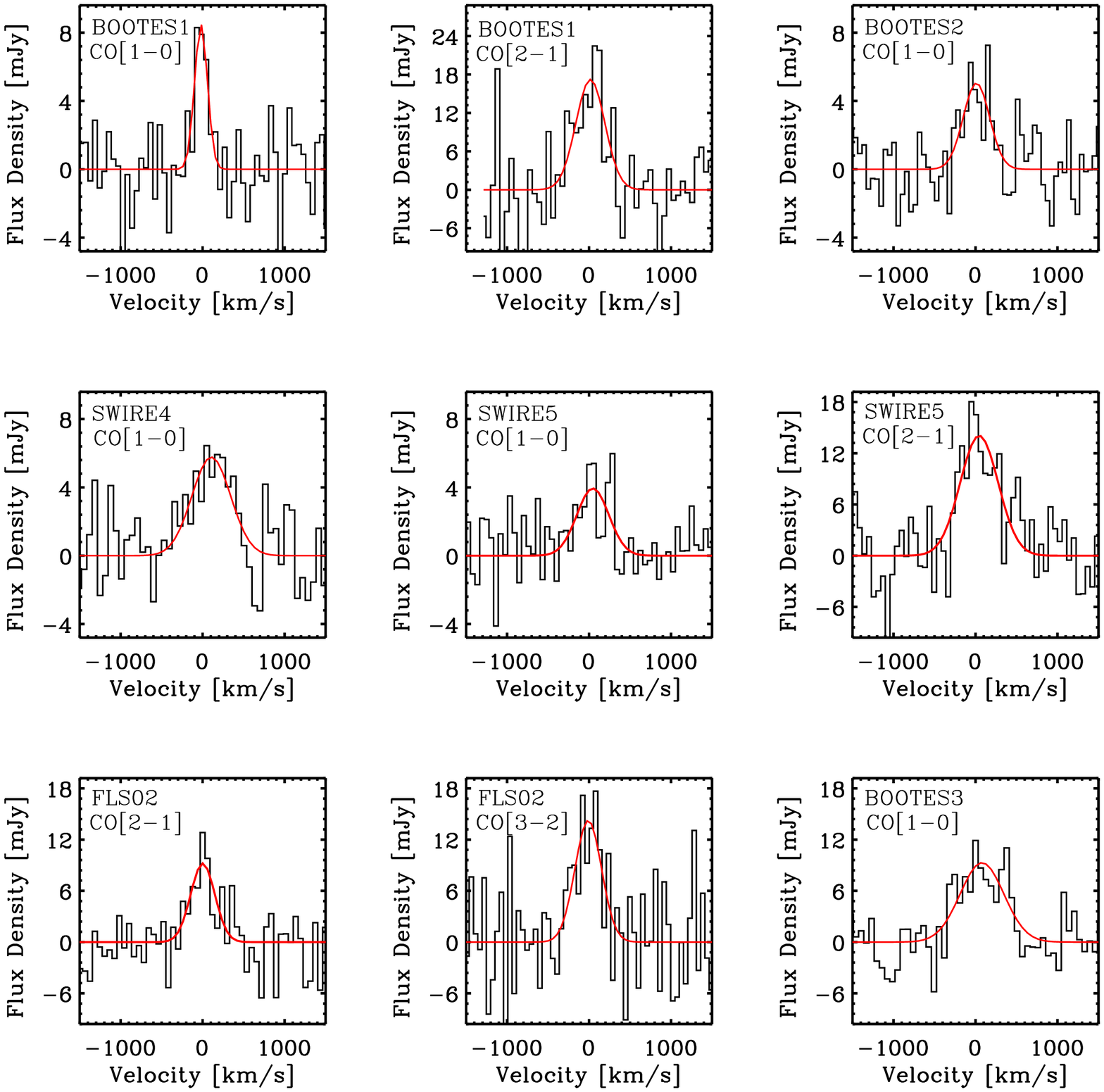}\\
\includegraphics[scale=0.36]{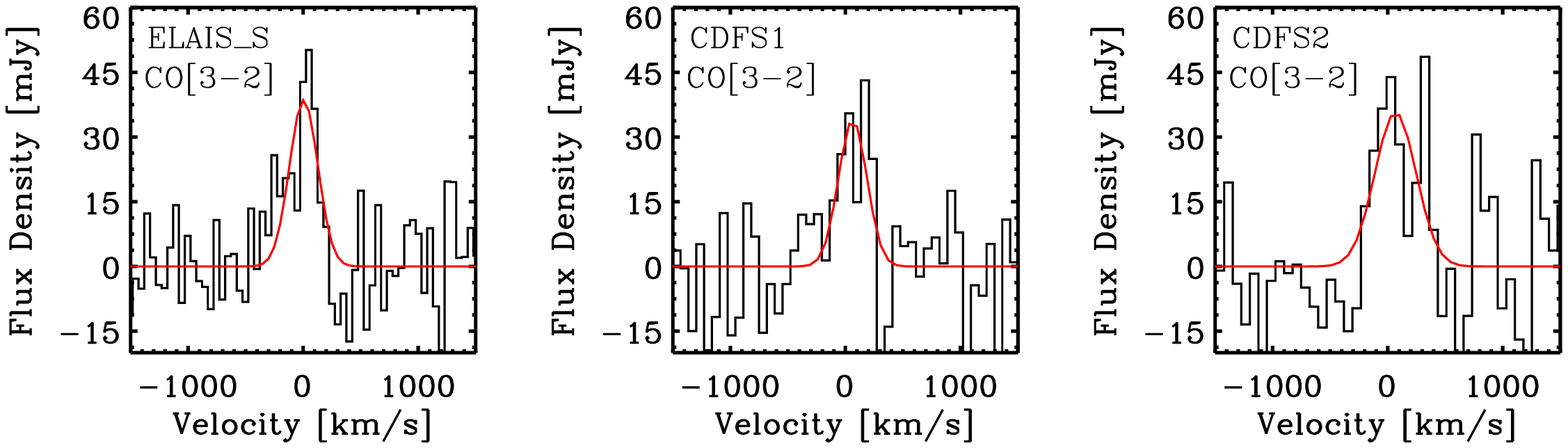}
\caption{CO spectra of the detected galaxies in 50$-$60\,km\,s$^{-1}$ bins. Zero velocity corresponds to the CO-determined redshift, listed in Table 4. According to the redshift, the lowest frequency observable is either the CO[1-0] or CO[2-1] or CO[3-2] line. The vertical scale is the flux density  in mJy. The last row depicts spectra obtained by APEX, while the rest were obtained by IRAM.}
\label{fig:co_spec} %
\end{figure}

We spent 2$-$5 hours on each galaxy, resulting in a noise level of 0.5$-$1.5\,mK per 30\,km\,s$^{-1}$ channel for all sources. The system temperatures were 
relatively stable with an average of $\sim$ 100\,K at 3\,mm, $\sim$ 280\,K at 2\,mm and $\sim$ 300\,K at 1.2\,mm, in $T^{*}_{\rm A}$. The pointing was regularly checked on continuum sources and yielded an accuracy of 3$''$ rms. The temperature scale used is in main beam temperature, $T_{\rm mb}$. At 3\,mm, 2\,mm and 1.2\,mm, the telescope half-power beam width is 27$''$, 17$''$ and 10$''$ respectively, and the main beam efficiencies ($\eta_{\rm mb}$ = $T^{*}_{\rm A}/T_{\rm mb}$) are 0.85, 0.70 and 0.64 respectively, with $S/T_{\rm mb}$ = 4.8\,Jy K$^{-1}$ for all bands. 

APEX observations were carried out in average weather conditions with precipitable water vapour 0.4 $<$ PWV $<$ 1.5 and in wobbler-switching mode, with a symmetrical azimuthal throw of 20$''$ and a frequency of 0.5\,Hz. Pointing was checked every hour and found to be better than 3$''$ (with a beam size of 20$''$). The focus was checked on the available planets, especially after sunrise when the telescope deformations are largest. The on source integration time was $\sim$ 2 hours for CDFS2 and ELAISS and $\sim$ 8 hours for CDFS1, targeting the CO[3$-$2] emission line in all sources using APEX1 and APEX2 (214 $-$ 275\,GHz and 267 $-$ 378\,GHz respectively). The achieved rms was around 0.4\,mK for CDFS1 and 0.9\,mK for the other two sources. 

Both sets of spectra (APEX and IRAM) were reduced using CLASS (within the GILDAS-IRAM package). In short, each spectrum was averaged and reduced using linear baselines, and then binned to 50$-$60\,km\,s$^{-1}$. To determine the properties of the emission lines (observed flux, velocity centroid and FWHM) we fitted Gaussian functions to the observed spectra. We obtained fits for all galaxies with a single Gaussian and with two Gaussian functions (in this case fixing the FWHM in each component to the same value). For all sources, a single Gaussian provided the best fit to the data, although in some cases there is evidence of  a double-horned profile. Considering a line detected when the integrated signal is larger than $>$3 $\sigma$, we have solid detections for at least one line per source. The median line-width of our sample is 416\,km\,s$^{-1}$ and ranges from 238 to 538\,km\,s$^{-1}$. Given the angular distance of the sources (average value 1000 Mpc), our beam subtends between 50 and 100 kpc, and all galaxies can be considered unresolved, at least as far as their molecular component is concerned. All spectra along with the best fit models are presented in Fig. \ref{fig:co_spec}  while the properties of the fitted lines are summarised in Table \ref{tab:co}.

Line luminosities, $L_{\rm CO}$ (measured in \lsol) and  \lco\ (measured in K\,km\,s$^{-1}$\,pc$^{2}$), were derived using equation (1) and (3) of Solomon \& Vanden Bout\,(2005): 

\begin{equation}
\begin{centering}
{L_{\rm CO} = 1.04 \times 10^{-3}~S_{\rm CO} \Delta\noindent u~\nu_{\rm rest}(1+z)^{-1}D^{2}_{\rm L}}
\end{centering}
\end{equation}
\noindent where $L_{\rm CO}$ is measured in \lsol, the luminosity distance $D_{\rm L}$ is in Mpc, the velocity integrated flux, $S_{\rm CO}\Delta$u, in Jy\,km\,s$^{-1}$, the rest-frame frequency of the observed line, $\nu_{\rm rest}$, in GHz, and 

\begin{equation}
\begin{centering}
{L^{\prime}_{\rm CO} = 3.25 \times 10^{7}~S_{\rm CO}\Delta\noindent u~\nu^{-2}_{\rm obs}D^{2}_{\rm L}(1+z)^{-3}}. 
\end{centering}
\end{equation}
\noindent As discussed, for four source  we targeted and detected the CO[1$-$0] emission line so the derived \lco\ corresponds to  $L^{\prime}_{\rm CO[1-0]}$. For the remaining 5 sources (FLS2, BOOTES2, CDFS1, CDFS2 and ELAISS) we converted the observed \lco\ = $L^{\prime}_{\rm CO[2-1]}$ or  \lco\ = $L^{\prime}_{\rm CO[3-2]}$, to $L^{\prime}_{\rm CO[1-0]}$ adopting $r_{\rm 21}$ = 0.75 and $r_{\rm 32}$ = 0.5 (e.g., Dannerbauer et al.\,2009, Ivison et al.\,2010, Rigopoulou et al.\, 2013, Aravena et al.\,2014). For XMM01 and HLSW-01 (SWIRE6) we adopt the CO[1-0] measurements reported by Fu et al\,(2013) and Scott et al.\,(2011) respectively.

\section{Derivation of FAR-IR Properties}
\begin{figure*}
\centering
\includegraphics[scale=0.35]{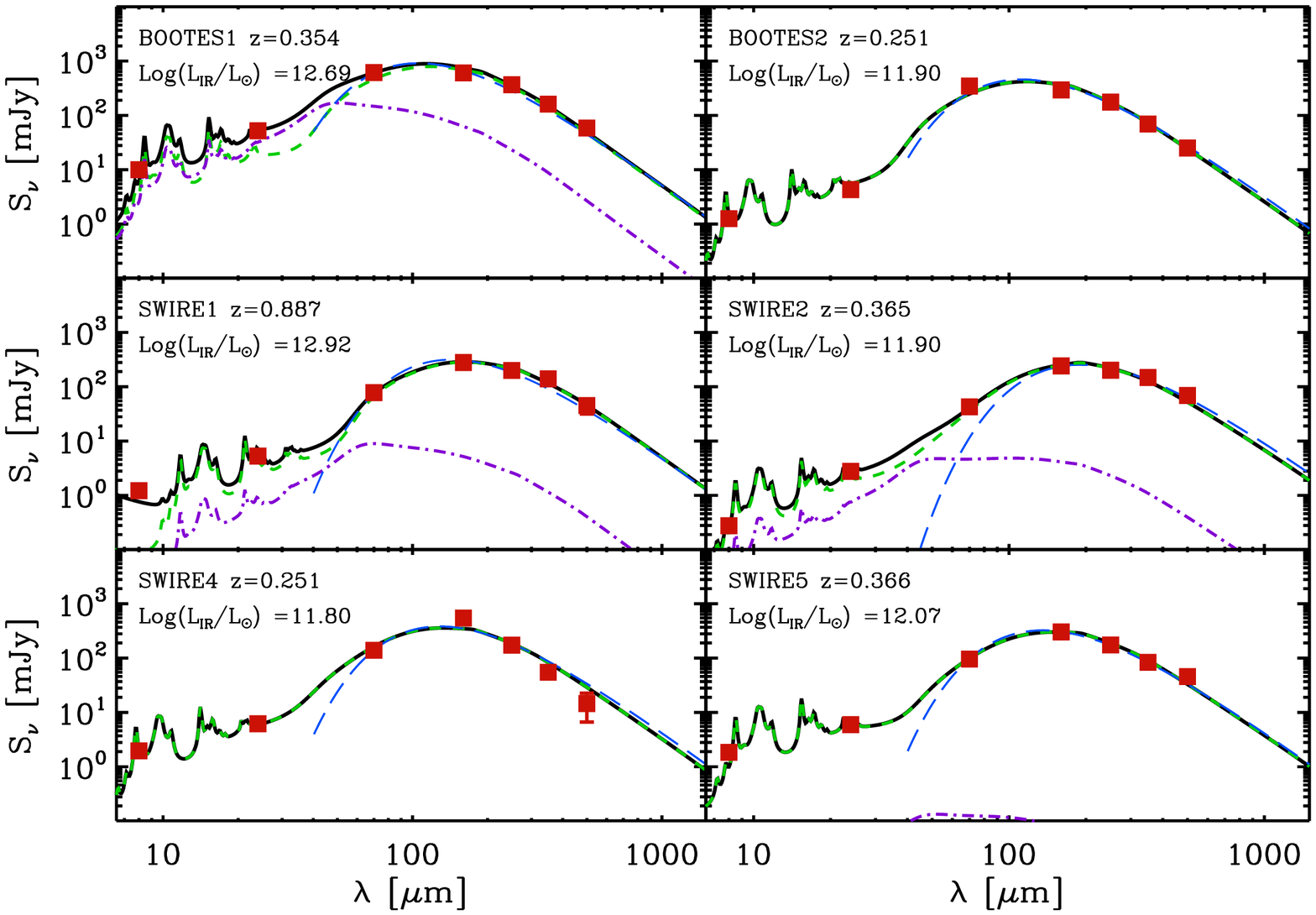}
\includegraphics[scale=0.35]{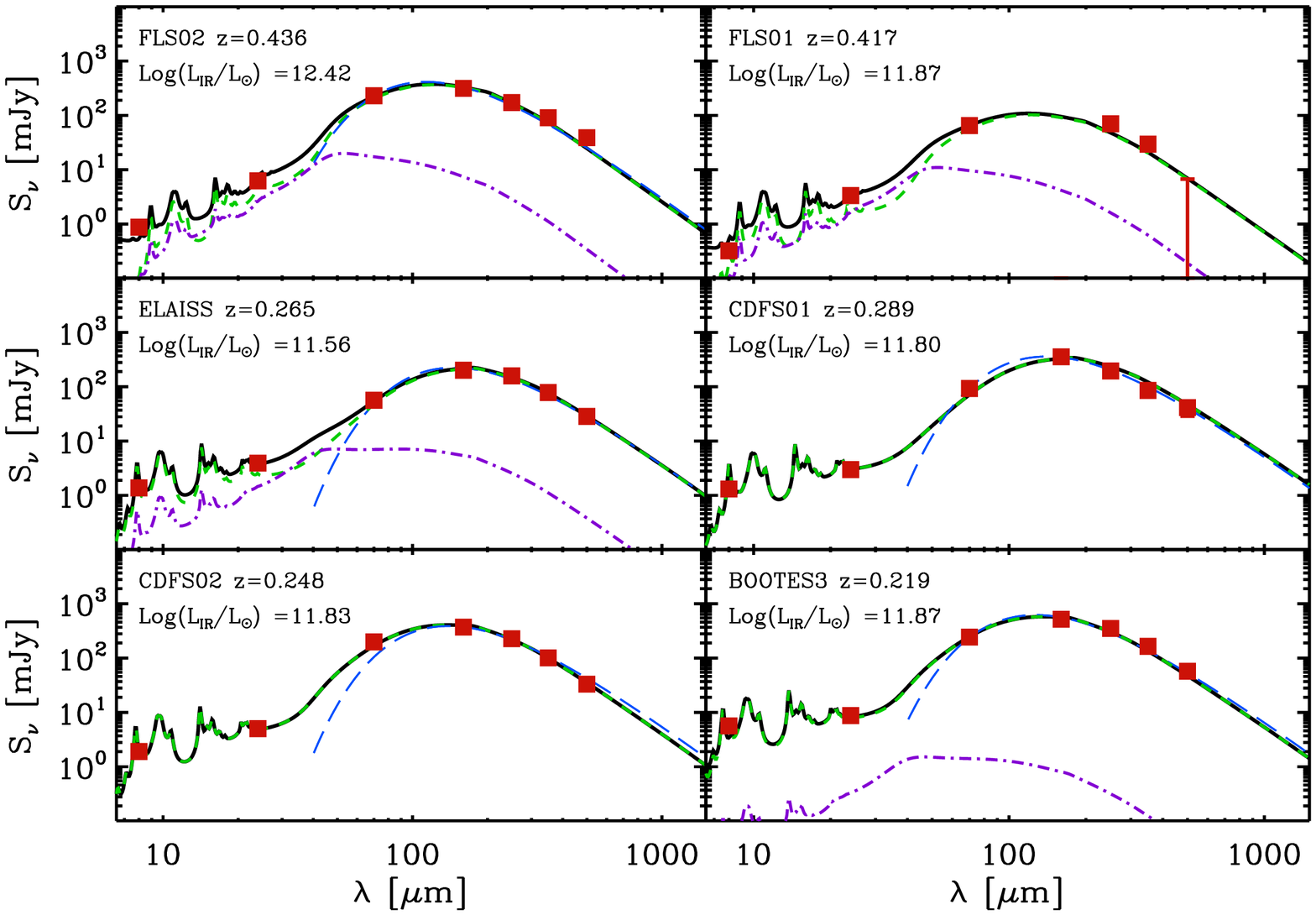}\\
\includegraphics[scale=0.35]{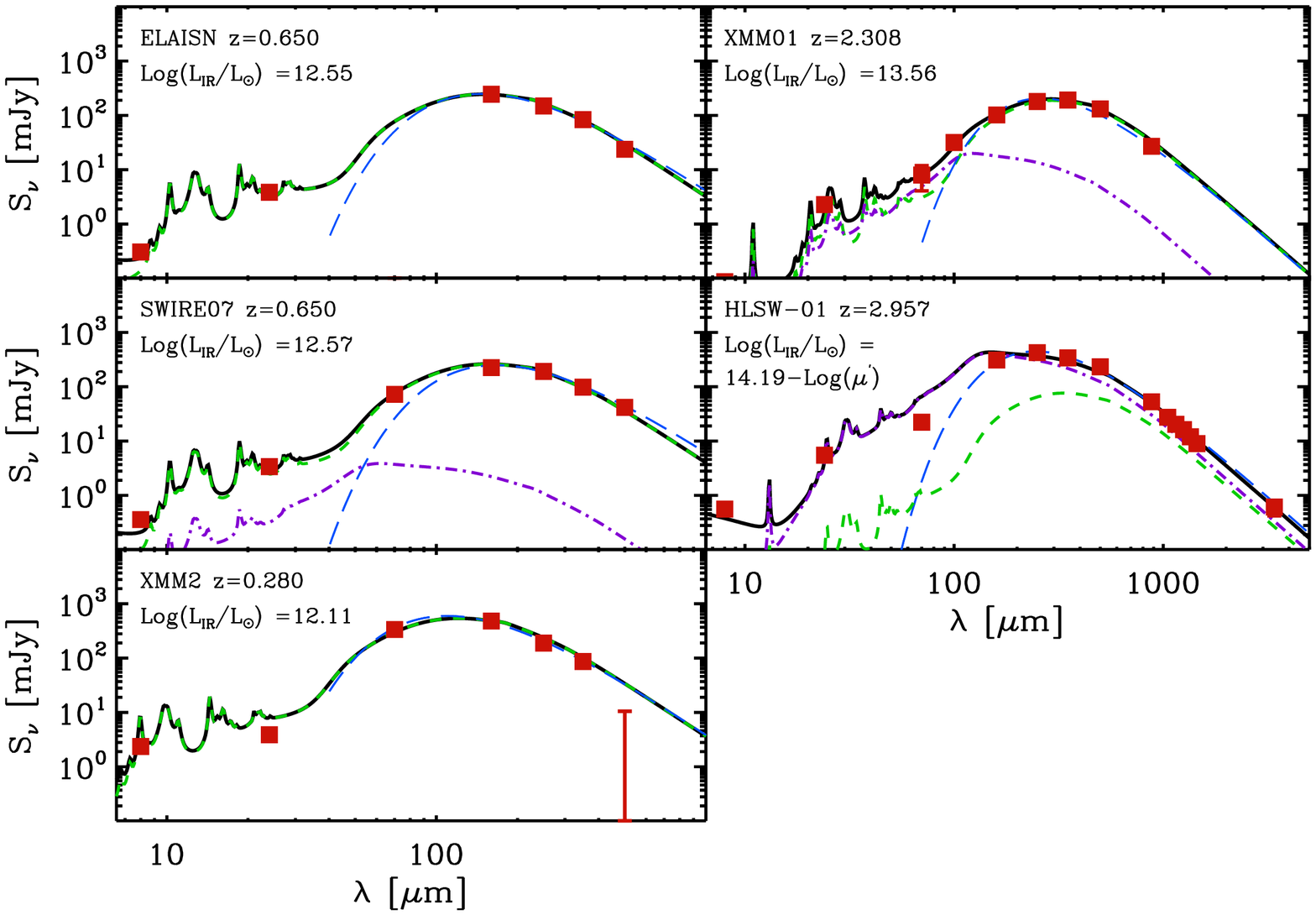}
\caption{Spectral energy distributions (observed frame) of sources from our sample, including IRAC 8$\,\mu$m, MIPS 24, 70 and 160$\,\mu$m (where available) and SPIRE 250, 350 and 500$\,\mu$m observed data points (red squares). The observed data are overlaid with the best-fit DL07 model (black line). The ``PDR''
and diffuse ISM components are shown in purple and green, respectively. The dashed blue line is the best-fit modified blackbody model with $\beta = 1.5$. For XMM01 and HLSW-01 we also include (sub)mm photometry from Fu et al.\,(2013) and Conley et al.\,(2011), respectively.}
\label{fig:irsed1} %
\end{figure*}

We combine ancillary \s\ IRAC (3.6$-$8.0\,$\mu$m), MIPS (24, 70, 160\,$\mu$m) photometry with new \h\ SPIRE observations at 250, 350 and 500$\,\mu$m, and use the Draine \& Li 2007 (DL07) model to derive dust masses and infrared luminosities by fitting the mid-IR to submm photometry. The analysis is also supplemented by a more simplistic but widely used single temperature modified black body (MBB) fit in order to derive a representative single dust temperature (\td) of the ISM. In particular we adopt a fixed effective dust emissivity index of $\beta$ = 1.5  and  fit observed data points with $\lambda_{\rm rest} >$ 50$\,\mu$m to avoid emission from very small grains that dominate at shorter wavelengths. A similar technique for the derivation of \md, \td\ and \lir\ using DL07 and MBB models is presented in detail in Magdis et al.\,(2012a) and Magdis et al.\,(2013). The best fit model SEDs are presented in Fig. \ref{fig:irsed1}  and the corresponding derived parameters are summarised in Table \ref{tab:dl07}. 

The infrared luminosities (\lir\ = $L{\rm[8-1000\,\mu m]}$) of the galaxies in the sample (excluding the two high$-z$lensed) vary between 3.8 $\times 
10^{11}\,$\lsol\  and 8.3 $\times 10^{12}\,$\lsol, with a median of 7.4 $\times 10^{11}$\,\lsol. Out of the whole $z < 1$ spectroscopic sample, 8 sources have  
1.0 $\times 10^{11} <$ \lir/\lsol\ $< 1.0 \times 10^{12}$ and 4 sources \lir/\lsol\ $> 1.0 \times 10^{12}$. Similarly, we probe a wide range of dust temperatures, \td\ = 29 $-$ 42\,K, and dust masses  \md\ $= 1.9 \times 10^{8} - 2.8 \times 10^{9}$\,\msol.

\begin{table*}
{\footnotesize
\caption{Far-IR Properties}             
\label{tab:dl07}      
\centering                         
\begin{tabular}{l c c c c}        
\hline\hline                 
Source&log $L_{\rm IR}$ $^{a}$ &log$L_{\rm FIR}$ $^{b}$& log $M_{\rm d}$ $^{c}$&$T_{\rm d}$ $^{d}$ \\
&[L$_{\odot}$]&[L$_{\odot}$]&M$_{\odot}$]&[K]\\
\hline
BOOTES1&12.69  $\pm$ 0.02 &      12.43  $\pm$ 0.04  &     9.06  $\pm$   0.16  &    42 $\pm$ 2\\
BOOTES2&11.91  $\pm$ 0.05  &     11.75 $\pm$  0.07   &    8.61 $\pm$     0.16   &   36 $\pm$ 2\\
BOOTES3&11.87  $\pm$ 0.04   &    11.66  $\pm$ 0.08    &   8.92   $\pm$  0.14    &  33 $\pm$ 2\\
CDFS1 & 11.79  $\pm$ 0.04   &    11.55 $\pm$  0.07    &   9.21  $\pm$   0.11    &  29 $\pm$ 2\\
CDFS2 &11.82  $\pm$ 0.03     &  11.63  $\pm$ 0.04     &  8.87   $\pm$  0.07    &  32 $\pm$ 2\\
ELAISS   & 11.59  $\pm$ 0.04   &    11.35 $\pm$  0.05    &   8.87  $\pm$   0.14   &   30 $\pm$ 2\\
FLS01     & 11.86  $\pm$ 0.03     &  11.66 $\pm$  0.05   &    8.29   $\pm$  0.14    &  39*\\
FLS02      & 12.41  $\pm$ 0.01     &  12.24 $\pm$  0.03   &    8.85  $\pm$   0.08    &  42 $\pm$ 2\\
SWIRE1   & 12.92 $\pm$  0.04  &     12.74 $\pm$  0.07   &    9.39   $\pm$  0.16    &  39 $\pm$ 2\\
SWIRE2   &11.90  $\pm$ 0.01   &    11.62  $\pm$ 0.02   &    9.46  $\pm$   0.09    &  28 $\pm$ 2\\
SWIRE4   &11.80  $\pm$ 0.01   &    11.60  $\pm$ 0.02   &    8.77  $\pm$   0.07   &   33 $\pm$ 3\\
SWIRE5   &12.06  $\pm$ 0.03   &    11.85  $\pm$ 0.06   &    9.07  $\pm$   0.10   &   33 $\pm$ 2 \\
HLSW-01$^{e}$ (SWIRE6)  & 13.15  $\pm$ 0.03    &   12.98 $\pm$  0.07   &    9.01  $\pm$   0.13   &   51 $\pm$ 2\\
XMM1       &  13.56  $\pm$ 0.03    &   13.36 $\pm$  0.04   &    10.07  $\pm$  0.23   &  39 $\pm$ 2\\
\hline
ELAISN    &   12.54 $\pm$  0.05 & 12.37 $\pm$  0.08   &    9.15 $\pm$    0.22  &    37 $\pm$ 2\\
SWIRE7   & 12.56 $\pm$  0.05    &   12.38  $\pm$ 0.08   &    9.27  $\pm$   0.18   &   36 $\pm$ 2\\
XMM2     & 12.11 $\pm$ 0.05     & 11.94 $\pm$ 0.09 & 8.83 $\pm$ 0.21 & 38 $\pm$ 2 \\
\hline           
\end{tabular}\\
Notes:\\
$^{a}$ \lir\ = $L\rm[8-1000\,\mu m]$.\\ 
$^{b}$ $L_{\rm FIR}$ = $L\rm[ 42.5-122.5\,\mu m]$.\\ 
$^{c}$ Dust masses as derived from Draine \& Li\,(2007) models. \\
$^{d}$ Dust temperatures as derived from modified blackbody fits with fixed $\beta = 1.5$ for sources with at least three photometric data points at $\lambda_{\rm rest}$ $>$ 50$\,\mu$m.  
$^{*}$ Based on the $L_{\rm 60}$/$L_{\rm 100}$ ratio from the best fit DL07 model.  
$^{e}$ Values corrected for magnification assuming $\mu = 10.9$(Gavazzi et al.\,2011)
}
\end{table*}

\begin{figure}
\centering
\includegraphics[scale=0.4]{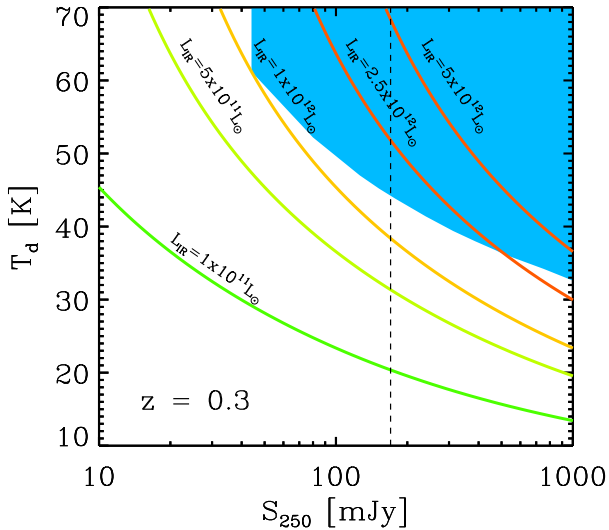}
\caption{ Dust temperature as a function of observed SPIRE 250$\,\mu$m flux density  as derived from a modified black body model with fixed $\beta = 1.5$ of a source at $z = 0.3$. The different colors corresponds to tracks of constant \lir, ranging from \lir\ = 1 $\times$ $10^{11}$\,\lsol\ (green line) to \lir\ = 5 $\times$ $10^{12}$\,\lsol\ (red line). The vertical line corresponds to $S_{\rm 250} = 170$\,mJy which is the selection criteria for our sample. The blue shaded region depicts the parameter space obtained from {\textit{IRAS}} observations at a detection threshold of $S_{\rm 60}$ = 0.4\,Jy.}
\label{fig:bias} %
\end{figure}

However, we note that by imposing a flux cut limit in the 250$\,\mu$m flux density we are actually selecting sources based on their cold dust 
emission (for the redshift range of our targets), which could introduce a selection bias towards colder sources. This is depicted in Fig. \ref{fig:bias}, where we show tracks of constant luminosity as a function of \td\ and $S_{\rm 250}$ for galaxies at $z$ = 0.3  using a MBB 
with fixed $\beta$ = 1.5. At this redshift, our criterion selects sources with \lir\ $\leq$ 1.0 $\times 10^{12}$\,\lsol\ and \td\ $\leq$ 40\,K, while missing warmer sources of similar \lir. On the other hand,  at higher luminosities (\lir\ $\geq$ 2.5 $\times$ $10^{12}\,$\lsol)  we only  miss extremely warm sources (\td\ $\geq$ 60\,K). To compare our selection to samples in the  pre-\h\ era we repeat the 
analysis, this time considering the {\textit{IRAS}} 60$\,\mu$m flux density. The blue shaded area in Fig. \ref{fig:bias}, shows the 
parameter space covered by {\textit{IRAS}} observations at a detection limit of 0.4\,Jy at 60$\,\mu$m (Faint Source Catalog, Moshir et al.\,1992). Our flux cut threshold would select the majority of the IRAS detected ULIRGs only missing  warm sources (\td\ $>$ 40\,K) in the narrow luminosity range of 1$-$2.5 $\times$ 10$^{12}$\,\lsol. However, given the sensitivity limitations of IRAS, at this redshift range the IRAS samples were predominantly limited to sources with \lir $>$ 3 $\times$ 10$^{12}$\lsol.  As an example all 0.2$ < z <$ 0.6, {\textit{IRAS}} selected ULIRGs presented by Combes et al.\, (2011), with 12.44 $<$ log(\lir/\lsol) $< $ 13.28 and 36.4\,K $<$ \td\ $<$ 56.6\,K would meet out our $S_{\rm 250}$ flux cut limit\footnote{None of 
these sources were selected in our sample as they were not covered by the \h\ HerMES observations.}.  Furthermore, Symeonidis et al.\,(2013), using both IRAS and \h\ selected samples report that the majority of (U)LIRGs (\lir\ $>$ 10$^{11}$\,\lsol) at all redshifts have mean dust temperatures  between 25 and 45\,K, with \td\ $<$ 25 and \td\ $>$ 45\,K sources being rare. As stated above, the  \td\ range of our sample is  29 $-$ 42\,K, very similar to the range of dust temperatures found by Symeonidis et al.\, (2013) for the bulk of the sources with infrared bolometric output comparable to that of the sources in our sample.
We therefore conclude, that our selection should be regarded fairly representative of the whole (U)LIRG population at the 0.2 $< z <$ 0.8 redshift range.

\section{The properties of the ISM of intermediate redshift (U)LIRGs} 

\subsection{The \lcii$-$\lir\ relation over cosmic time} 

Various Galactic and extragalactic studies of the [CII] emission (e.g., Tielens \& Hollenbach 1985, Stacey et al.\,1991, Meijerink et al.\,2007, Rigopoulou et al.\,2013) have established that it predominantly originates from PDRs on the outer layers of molecular clouds exposed to intense FUV radiation and have put it forward as a potentially  powerful tracer of the star formation activity. Indeed, using \lfir\footnote{($L_{\rm FIR}$ = $L_{42.5-122.5\,\mu m}$) as a star formation rate indicator (e.g., Kennicutt et al.\,1998), the majority of normal galaxies in the local Universe seem to obey a universal linear  \lcii$-$\lfir\   relation with a constant \lcii/\lfir\ ratio, albeit with a significant scatter (e.g., Malhotra et al.\,2001)}. However, extending the study of [CII] in local (U)LIRGs has revealed that they appear to deviate from this relation exhibiting lower \lcii/\lfir\ ratios, i.e., they are \textit{[CII]-deficient} with respect to their infrared luminosity (e.g., Malhotra et al.\,1997, Luhman et al.\,1998,2003, Diaz-Santos et al.\,2013, Farrah et al.\,2013). 

Extrapolating observations of the local Universe to high$-z$ galaxies, a natural expectation would be that high$-z$ (U)LIRGs would also appear [CII]-deficient with respect to local normal galaxies. However, the detections of [CII] emission from distant ($z >$ 1), star forming galaxies have come to challenge this picture, revealing that  ULIRGs can exhibit \lcii/\lfir\ ratios similar to that of local normal galaxies and a factor of $\sim$10 larger compared to local (U)LIRGs (e.g., Hailey-Dunsheath et al.\,2010, Stacey et al.\,2010, Valtchanov et al.\,2011, Riechers et al.\,2013). While these studies seem to suggest a strong evolution in the [CII] emission between local (\textbf{$z <$ 0.2}) and $z >$ 1 (U)LIRGs, it is unclear what drove this evolution.

Rigopoulou et al.\,(2014), found that the majority of \h$-$selected (U)LIRGs have \lcii/\lfir\ ratios similar to that of high$-z$ star formation dominated ULIRGs and local normal galaxies, suggesting that the transition of the properties in the ISM of the (U)LIRG phenomenon was already in place by $z \sim 0.3$. Here, by using the same sample, we bridge the gap between the local and high$-z$ galaxies and provide a continuous sampling of the \lcii$-$\lfir\ relation over the last 10 billion years for a wide range of extragalactic sources.
\begin{figure*}
\centering
\includegraphics[scale=0.6]{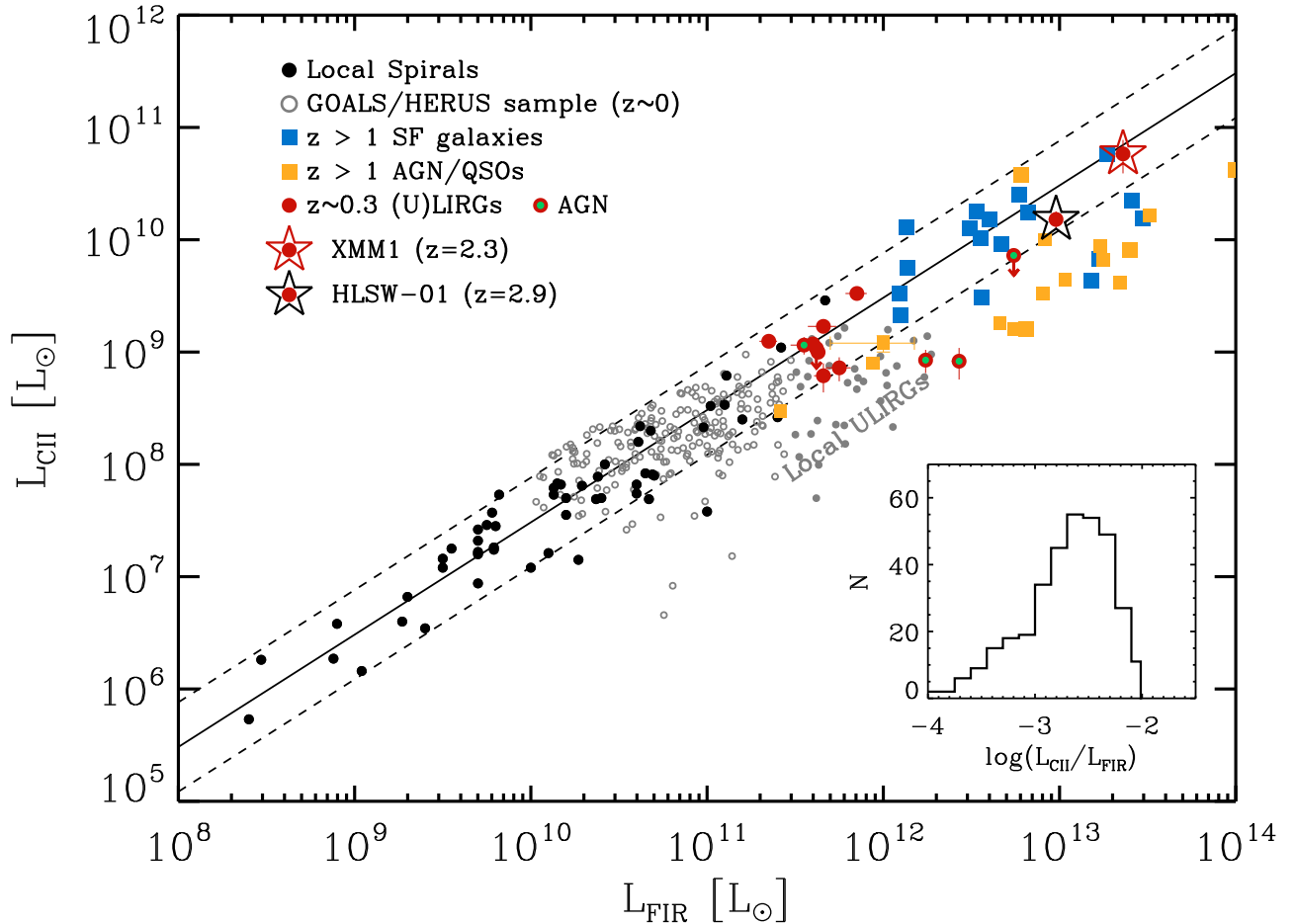}
\caption{[CII] emission line luminosity (\lcii) versus far infrared (42.5$-$122.5$\,\mu$m) luminosity (\lfir). Our sample of intermediate redshift (U)LIRGs is shown as red circles. Sources from our sample with an AGN are indicated with a green filled circle, undetected sources are downward arrows (3\,$\sigma$ upper limits) and  XMM01 and HLSW-01 as a large stars. The rest of the data are: local normal galaxies (filled black circles, from Malhotra et al.\,2001); local (U)LIRGs from the GOALS and HERUSS sample (empty grey circles from DIaz-Santos et al.\,2013 and Farrah et al.\,2013)' high$-z$ ($z > 1$) star-forming galaxies (blue squares, from Stacey et al.\,2010, Hailey-Dunsheath et al.\,2010, Valtchanov et al.\,2010, Ivison et al 2010, Cox et al.\,2011, De Breuck et al.\,2011, Swinbank et al.\,2012, George et al.\,2013, Riechers et al.\,2013, Rawle et al.\,2014,  Ferkinhoff et al.\,2014, Riechers et al.\,2014); and high$-$z AGN dominated sources and QSOs (filled orange squares, from Iono et al.\,2006, Walter et al.\,2009, Wagg et al.\,2010, Gallerani et al.\,2012, Wang et al.\,2012, Venemans et al.\,2012, Carilli et al.\,2013, Willott et al.\,2013). Gray filled circles correspond to local (U)LIRGs with $L_{IR}$ $>$ 5 $\times$ 10$^{11}$,\lsol. The solid black lines correspond to the best linear fit to the local normal galaxies of Malhotra et al.\,(2001) with a scatter of 0.3\,dex (dashed lines) and a slope of unity. The inset panel shows the distribution of the \lcii/\lfir\ ratio for the all sources included in this plot.}
\label{fig:lciilir} %
\end{figure*}
In Fig. \ref{fig:lciilir} we plot the [CII] line luminosity as a function of far-infrared luminosity for: local normal galaxies (Malhotra et al.\,2001); local (U)LIRGs from the GOALS and HERUS samples (Diaz-Santos et al.\,2013, Farrah et al.\,2013); high$-z$ ($z > 1$) star forming galaxies (Stacey et al.\,2010, Hailey-Dunsheath et al.\,2010, Valtchanov et al.\,2010, Cox et al.\,2011, Ivison et al 2010, Swinbank et al.\,2012, Rawle et al.\,2013, George et al.\,2013, Riechers et al.\,2013, Ferkinhoff et al.\,2014); high$-z$ AGNs and QSOs (Iono et al.\,2006, Wagg et al.\,2010, Gallerani et al.\,2012)  and for our sample of $z \sim 0.3$ (ULIRGs). Fitting the local normal galaxies yields:
\begin{equation}
\begin{centering}
{\rm{log}({\textit{L}}_{\rm CII}) = (-2.51 \pm 0.39) + \rm{log}({\textit{L}}_{\rm FIR})}, 
\end{centering}
\end{equation}
 revealing a constant \lcii/\lfir\ ratio  albeit with a considerable scatter. As shown in Figure \ref{fig:lciilir}, the majority of star forming galaxies (with [CII] measurements) at all redshifts and infrared luminosities follow this relation (within the scatter), suggesting that \lcii\ can trace \lfir, and therefore the  star formation rate, within a factor of 2.5 at all epochs. However, as discussed above and as  evident  from the skewed tail of the \lcii/\lfir\ distribution (inset panel of  Fig. \ref{fig:lciilir}) outliers from this relation do exists and are predominantly the local (U)LIRGs (\lfir\ $>$ 5 $\times$ 10$^{11}$\,\lsol), high$-z$ AGN dominated sources, and a small fraction of high$-z$ star forming galaxies. These sources appear \textit{[CII]-deficient} with respect to their \lfir\ exhibiting, on average, a lower \lcii/\lfir\ ratio compared to that of local normal galaxies and the majority of high$-z$ star formation dominated  ULIRGs. This trend implies an evolution in the star forming regions  of (U)LIRGs. Indeed, focusing on our sample of intermediate redshift (U)LIRGs, we find that only two out of the 14 targets appear to be \textit{[CII]-deficient} while the rest exhibit \lcii/\lfir\ ratios similar to that of local normal galaxies and high$-z$ star forming galaxies. Interestingly, these two sources are also classified as AGN. In what follows we will investigate the impact of an AGN on the observed \lcii/\lfir\ ratio of our sample.
\begin{figure*}
\centering
\includegraphics[scale=0.7]{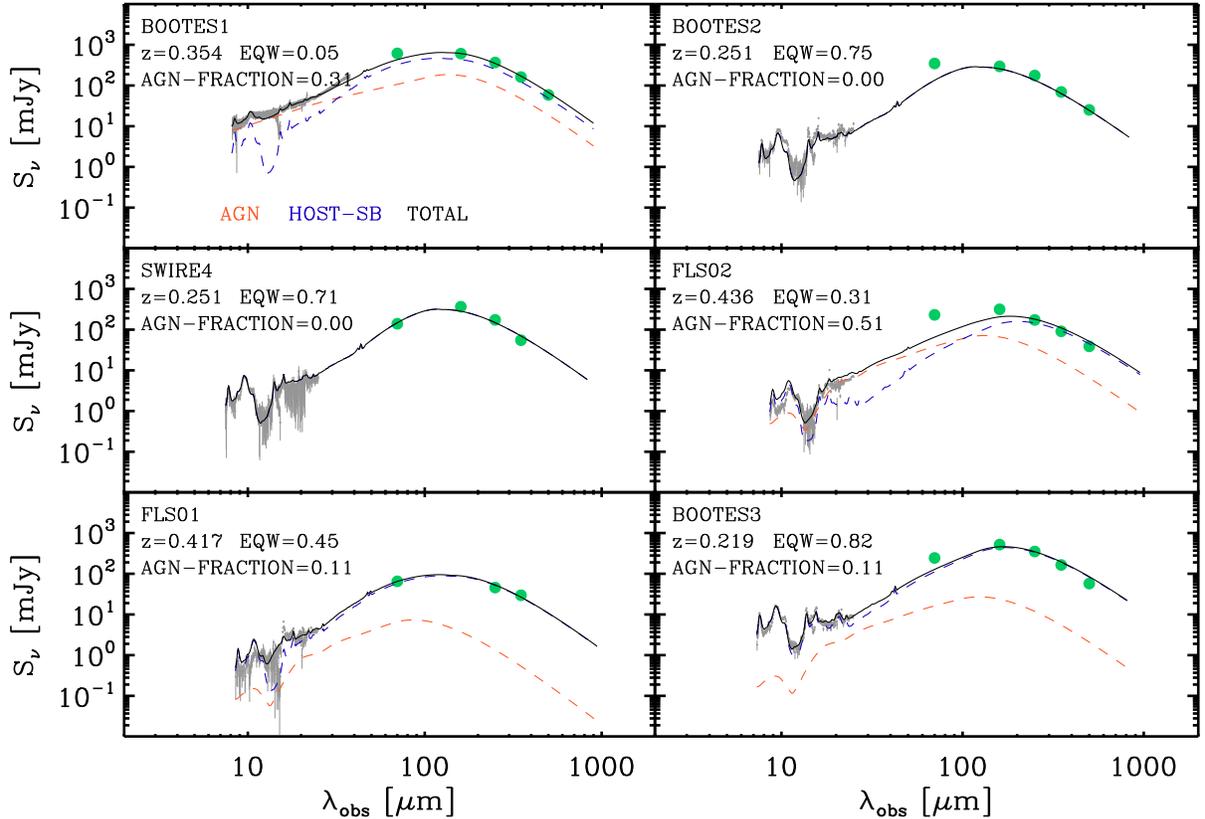}

\caption{SED fitting using the AGN/host-star-forming galaxy decomposition technique of Mullaney et al.\,(2011) for sources in our sample with available mid$-$IR IRS spectroscopy (grey points). Green circles show the observed far-IR photometric points and are overlaid with the best-fit total model (black line). The AGN and host galaxy components are shown as red and blue dashed lines, respectively. The quoted EQW corresponds to the equivalent width of the 6.2$\mu$m PAH feature.}
\label{fig:sed_agn} %
\end{figure*}
\subsection{The effect of an AGN}
A possible explanation for deviations from the general \lcii$-$\lfir\ trend could be the a strong contribution of an AGN to the bolometric infrared output of the source.  If \lcii\ is predominantly associated to star formation while \lfir\ arises from a mixture of AGN and star formation activity, that would result in a decreased \lcii/\lfir\ ratio with respect to purely star-forming galaxies. This scenario holds under the assumption that even if [CII] can be excited by AGN activity (both in PDRs and XDRs), its contribution to the total [CII] emission is typically found to be small (e.g., Crawford et al.\,1985, Unger et al.\,2000, Stacey et al.\,2010). 

To investigate this scenario we perform 
an AGN/host-galaxy decomposition for five sources in our sample that benefit from mid-IR IRS spectra, using the SED decomposition method of 
Mullaney et al.\,(2011). This method employs a host-galaxy and an intrinsic AGN template SED to measure 
the contribution to the infrared output of these two components. The technique identifies the best-fitting model SED for the observed infrared 
data (spectra and photometry) through $\chi^{2}$ minimization and by varying the values of a set of free parameters (for details see Magdis et al.\,2013). The best-fit model SEDs are shown in Fig.  \ref{fig:sed_agn}. Out of the five sources with available IRS spectra two are classified as AGNs (Bootes1 and FLS2) and three as star-forming or composites based on optical or mid-IR spectroscopy. 

From the various output parameters  we focused on the contribution of an AGN in the far-infrared output ($f_{\rm AGN}$). For star-forming and composite sources (classified based on the equivalent width of the 6.2$\,\mu$m PAH feature) we find a negligible (if any) contribution of an AGN to \lfir. For the two AGN, which also happen to have \lcii/\lfir\ similar to that of local ULIRGs, we derive $f_{\rm AGN} = 0.31$ (BOOTES1) and $f_{\rm AGN} = 0.51$ (FLS2). We note that given the poor fit to the rest frame mid-IR data for these two sources, these values should be treated with caution and  the exact contribution of an AGN to the \lfir\ remains largely unconstrained. However, even if we correct the \lfir\ by a factor of 2,  the  \lcii/\lfir\ of these two sources remain approximately $\sim$3 times lower compared to the rest of the sample. Therefore, while the existence of an AGN can partially affect the \lcii/\lfir\ values for these systems, the AGN contribution to \lfir\ cannot fully explain the observed [CII] deficit. Furthermore, the rest of the sources in our sample that are classified as AGN fall within the scatter of the \lcii\ $-$ \lfir\ relation defined in equation 3, suggesting that there is no direct correlation between the existence of an AGN and the observed \lcii/\lfir\ ratio. Similar results have been reach by Diaz-Santos et al.\,(2013) and  Farrah et al.\,(2013) based on a sample of local ULIRGs and using a number of far-IR lines. However, we should stress that in our sample, as well as in the majority of local ULIRGs, the far-IR output is dominated by star formation (e.g., Genzel et al.\,1998, Rigopoulou et al.\,1999, Armus et al.\,2007, Desai et al.\,2007, Petric et al.\,2011). On the other hand, sources where AGN activity dominates the far-IR output of the sources, both in the local Universe as well as at high$-z$,  the origin of the [CII]-deficit can be explained through the contamination of \lfir\ by AGN activity  (e.g., Stacey et al.\,2010, Wagg et al.\,2010, Sargsyan et al.\,2012).

Therefore, we cannot rule out that for sources where AGN activity dominates the far-IR output of the sources, both in the local Universe as well as at high$-z$,  the origin of the [CII]-deficit as reported in various studies is not due to contamination of \lir\ by AGN activity  (e.g., Stacey et al.\,2010, Wagg et al.\,2010, Sargsyan et al.\,2012).
\begin{figure*}
\centering
\includegraphics[scale=0.28]{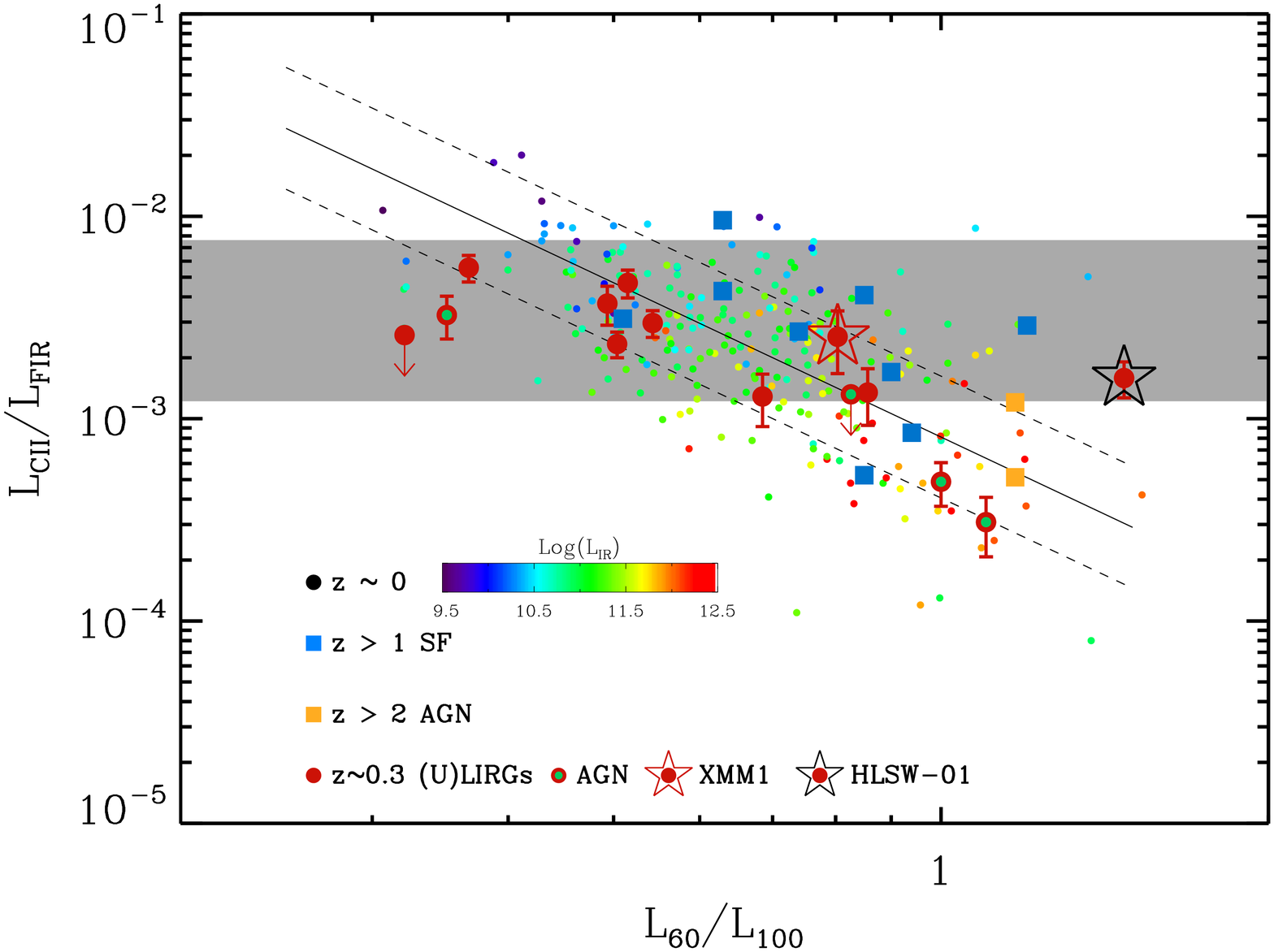}
\includegraphics[scale=0.28]{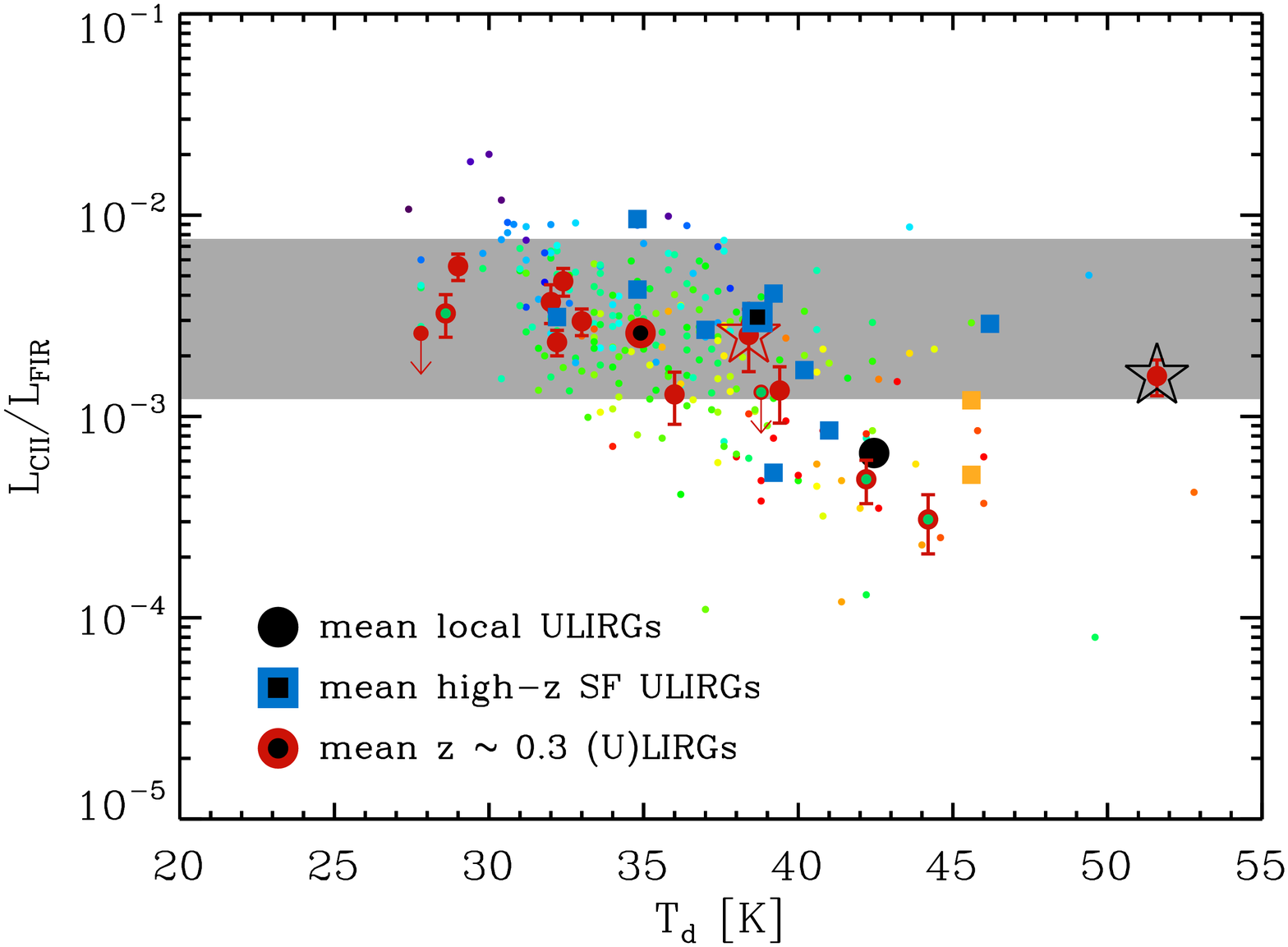}
\caption{{\textbf {Left:}} Ratio of [CII] line luminosity to far-infrared luminosity (\lfir) as a function of the
$L_{\rm 60}$/$L_{\rm 100}$ continuum luminosity ratio for local (U)LIRGs (GOALS and HERUS), intermediate redshift (U)LIRGs, high$-z$ star forming galaxies, and high$-z$ QSOs. The local sample is color coded based on the luminosity of the sources. The  plot reveals  a decrease of the \lcii/\lfir ratio for warmer sources at all redshifts. The solid line corresponds
to a linear fit of the all data in log-log space. The parameters of the fit are given in eq.\,(6). The dotted lines correspond to the $\pm$1\,$\sigma$ uncertainty of the derived relation.  The vertical shaded region covers the range of \lcii/\lfir\ values within 1\,$\sigma$ of eq.\,(3).  {\textbf {Right:}} Same as the left panel but using a modified black body with fixed $\beta = 1.5$ and the redshift of the sources to convert the observed continuum luminosity ratio to dust temperature (\td). We also include the mean values for the local, intermediate redshift and high-z ULIRGs.}
\label{fig:color} %
\end{figure*}
 \subsection{\lcii$-$\lfir\ as a function of \td}
 Various studies of [CII] emission from normal galaxies and local (U)LIRGs have established an observational trend where sources with lower \lcii/\lfir\ values tend to have warmer dust temperatures (e.g., Malhotra et al.\,1997). Since the \lcii/\lfir\ ratio depends on the efficiency of the heating of gas by photoelectrons from dust, it is indeed expected to be sensitive to the physical conditions of the ISM. A possible explanation for the observed trend was offered by Malhotra et al.\,(1997) based on the study of the colors and the [CII] emission of local ULIRGs. They suggested that the [CII]-deficit of warmer sources could be explained by a high $G_{\rm 0}/n_{\rm H}$ ratio in the neutral galactic medium, where $G_{\rm 0}$ is the incident far-UV 6$-$13.6\ eV radiation field and $n_{\rm H}$ the total hydrogen gas density. In this scenario, more extreme $G_{\rm 0}/n_{\rm H}$ ratios would lead to an increase in the positive charge of the dust particles, which  would result in a reduction of the amount of photo-electrons (with sufficient kinetic energy) released from dust grains, and consequently, in a decrease of the efficiency in the transformation of the incident UV radiation into gas heating.
  
 In a more general scenario, the observed deficits of various far-IR atomic lines could be the result of a higher average ionization parameter of the ISM, $\langle U \rangle$ (Luhman et al.\,2003, Abel et al 2009), defined as the dimensionless ratio of the incident ionising photon density to the hydrogen density: 
 
\begin{equation}
\begin{centering}
{U = \frac{Q_{\rm H}}{4 \pi R^{2}n_{\rm H}c}},  
\end{centering}
\end{equation}  
where $Q_{\rm H}$ is the number of hydrogen ionizing photons, $R$ is the distance of the ionizing sources to the PDR, and c is the speed of light.  
Assuming an average stellar population and size of star-forming regions, the above relation yields :
\begin{equation}
\begin{centering}
{U \propto \frac{G_{\rm 0}}{n_{\rm H}}}. 
\end{centering}
\end{equation}  
For fixed $n_{\rm H}$, higher levels of ionization (i.e., higher $\langle U \rangle$ values), would result in a higher ratio of ionized to atomic hydrogen and therefore in a decreased gas opacity in the HII regions, since less hydrogen atoms will be available to absorb UV photons. Subsequently, this will lead to a decreased gas-to-dust opacity, as dust will have to compete for UV photons with fewer neutral hydrogen atoms in the HII regions. As a consequence, a significant fraction of the UV radiation is eventually absorbed by large dust grains before being able to reach the neutral gas in the PDRs and
ionize the PAH molecules (Voit\,1992, Gonzalez-Alfonso et al.\, 2004, Abel et al.\, 2009), causing a deficit of photo-electrons and
hence, suppression of the [CII] line with respect to the total FIR dust emission. Higher $\langle U \rangle$ values will also translate into larger numbers of photons per dust particle and therefore higher dust temperature or warmer far-IR colors for sources with lower \lcii/\lfir\ (or other atomic lines) ratios. We note that the assumption of fixed $n_{\rm H}$ appears to be valid, based on the observation of Malhotra et al.\,(2003) that ($L_{\rm OI}+L_{\rm CII})/$\lfir\ varies as a function of far-IR color ($S_{\rm 60}/S_{\rm 100}$).

In Fig. \ref{fig:color} we plot the \lcii/\lfir\ ratio as a function of the $L_{\rm 60}/L_{\rm 100}$ color and dust temperature, for the same sample of galaxies as in Fig \ref{fig:lciilir}. For the local galaxies we used the available {\textit{IRAS}} photometry, while for our intermediate$-z$ and high$-z$ literature sources  we convolved the best-fit SED with the {\textit{IRAS}} filters. To convert the observed flux ratios of the sources drawn form the literature to dust temperatures we assumed a modified blackbody with a fixed dust emissivity index $\beta=1.5$. In both panels the local sample is color-coded, based on the infrared luminosity of the sources. It appears that the strong anti-correlation between the \lcii/\lfir\ ratio and the dust temperature, as already observed in the local Universe by various authors (e.g., Malhotra et al.\,2003, Diaz-Santos et al.\,2013), holds at all luminosities and all redshifts. The best fit to the data yields the  relation 

\begin{equation}
\begin{centering}
\rm log\left(\frac{{\textit{L}}_{\rm CII}}{{\textit{L}}_{\rm FIR}}\right) = (-3.09 \pm 0.06) + (-2.53 \pm 0.21) log\left(\frac{{\textit{L}}_{\rm 60}}{{\textit{L}}_{\rm 100}}\right)
\end{centering}
\end{equation}

\noindent with a scatter of 0.3\,dex, suggesting a strong dependence between the \lcii/\lfir\ ratio and the dust temperature at all epochs, with warmer sources exhibiting lower \lcii/\lfir\ ratios. 

This trend also provides hints about the physical origin of the observed difference of the \lcii/\lfir\ ratio between the local and distant ULIRGs. Several studies have 
shown that the majority of high$-z$ star-forming galaxies with ULIRG-like luminosities are colder than galaxies in the local Universe 
with similar \lir\ (e.g., Hwang et al.\,2010, Magdis et al.\,2010,2012, Symeonidis et al.\,2013). This is clearly depicted in Fig. \ref{fig:color}, 
where the bulk of high$-z$ star forming galaxies from the literature that have 
[CII] measurements, exhibit lower $L_{\rm 60}/L_{\rm 100}$ with respect to 
local ULIRGs. Note that all high$-z$ star-forming galaxies in this sample 
have \lir\ $>$ 10$^{12}$\,\lsol. Similarly, we find that our sample of \h-
selected intermediate redshift (U)LIRGs are on average colder than local 
ULIRGs with $\langle$\td$\rangle$ = 34\,K compared to 42\,K for local 
ULIRGs. 

As discussed above, the observed trend is compatible with the physical scenario in which the variation of the \lcii/\lfir\ is caused by an increase of $\langle U \rangle$ in [CII]-deficient sources. Diaz-Santos et al.\,(2013),\,(2014), showed that for local (U)LIRGs the observed decrease of  \lcii/\lfir\  ratio in warmer sources, i.e., sources with higher $\langle U \rangle$, is linked to the luminosity surface
density of the mid-IR emitting region;  warm, [CII]-deficient sources are associated with compact starbursts. Furthermore, based on resolved [CII] observations, they found that [CII]-deficient regions of local luminous LIRGs are restricted to their nuclei with extra nuclear 
emission showing  \lcii/\lfir\  ratios similar to that of normal galaxies. The implications of these findings for the high$-z$ galaxies are discussed in Section 6.

Finally, we note that variation of the \lcii/\lfir\ ratio could also be driven by galactic scale metallicity effects.  For example, higher dust temperatures could result from the increased hardness of the radiation field  and the transparency of the ISM towards lower metallicities, which enhances the ionization of gas. In combination with the longer mean free path lengths of ionizing photons in metal-poor galaxies, the filling factors of ionized gas media are bound to grow drastically (e.g., De Looze et al.\,2014). Furthermore, more UV photons might,  reach the neutral PDR gas in low metal abundance environments, where the hard radiation field reduces the PAH abundance and/or charges dust grains in PDRs, resulting in a reduced [CII] line emission. Apart from the CII-deficit, metallicity variations could also be responsible for the observed scatter in the \lcii-\lfir\ relation.
\begin{figure*}
\centering
\includegraphics[scale=0.45]{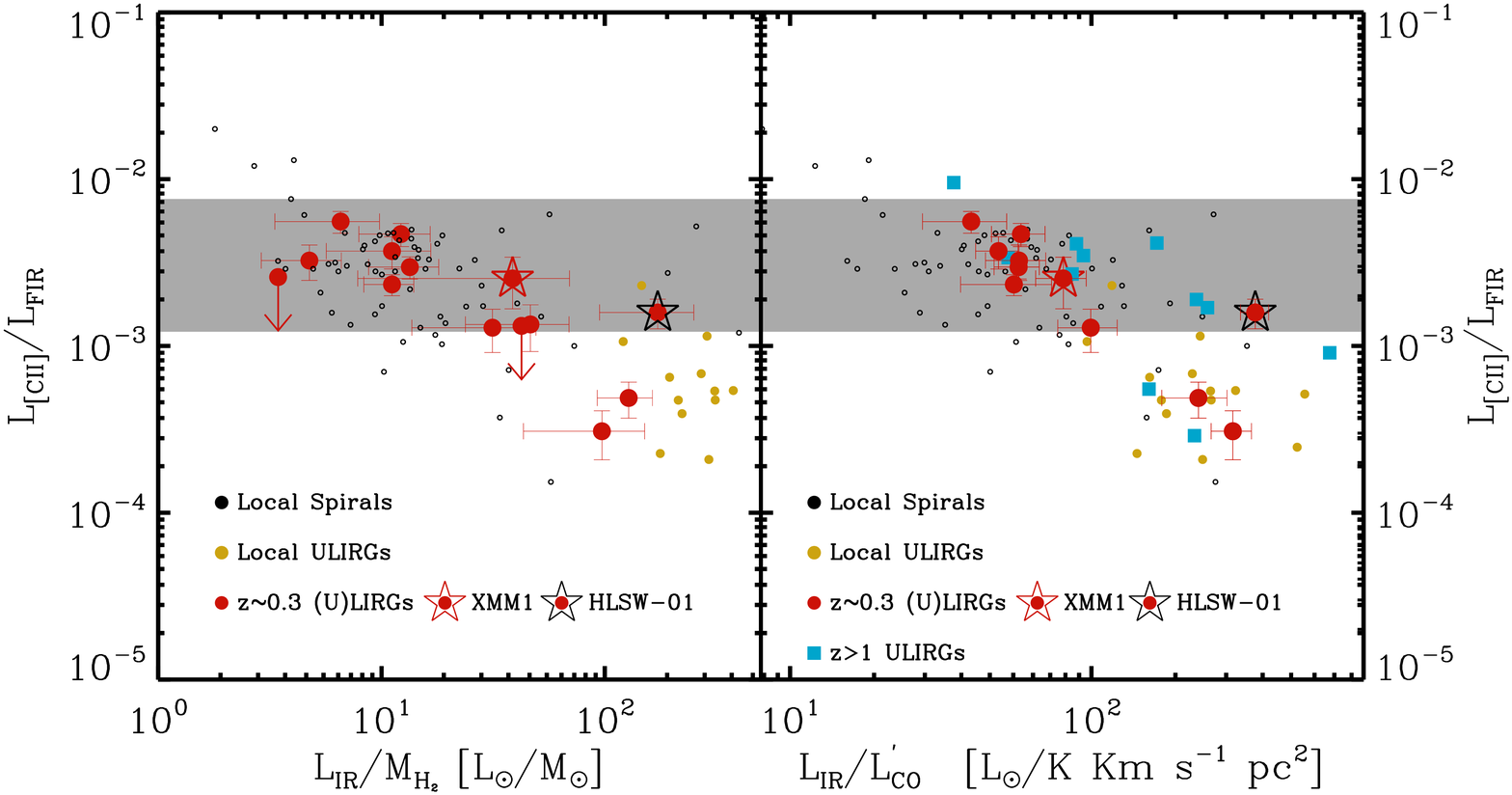}
\caption{ {\textbf {Left:}} \lcii/\lfir versus \lir/\Mmol for all (U)LIRGs is our sample that have [CII] observations. The vertical shaded region covers the range of \lcii/\lfir\ values within 1\,$\sigma$ of eq.\,(3).  {\textbf {Right:}} \lcii/\lfir versus \lir/\lco for all (U)LIRGs is our sample that have [CII] and CO observations.}
\label{fig:gas} %
\end{figure*}

\subsection{\lcii$-$\lir\ and star formation efficiency}
In a recent study, Gracia-Carpio et al.\,(2011), reported a strong anti-corellation between the relative 
strength of the far-IR lines and the infrared luminosity to molecular gas mass ratio, \lir/$M_{\rm H_{\rm 2}}$. In particular they argued that galaxies with  high star formation efficiencies (\lir/$M_{\rm H_{\rm 2}}$, Sanders \& Mirabel 1985), tend to have weaker far-IR emission lines as a consequence of an increased ionization parameter. Indeed, in a simplified scenario, the radiation field should be proportional to the number of available ionising photons (\lir) per hydrogen molecule ($M_{\rm H_{\rm 2}}$). They concluded that for fixed hydrogen density the observed [CII] deficit that starts to appear at \lir/$M_{\rm H_{\rm 2}}$ $>$ 80\,\lsol/M$_{\odot}$ can be understood if the ionisation parameter is an order of magnitude larger compared to sources with lower  \lir/$M_{\rm H_{\rm 2}}$ ($\langle U \rangle$ $\sim$ 10$^{-3}$ to 10$^{-2.5}$, for the latter). Interestingly the value of  \lir/$M_{\rm H_{\rm 2}}$ $\approx $ 80\,\lsol/M$_{\odot}$ where the far-IR  line deficits start to manifest is also similar to the limit that separates between the two modes of star formation (merger driven starbursts vs normal quiescent galaxies, e.g., Daddi et al.\, 2010, Genzel et al.\, 2010). Here, we test this scenario for our sample of intermediate redshift (U)LIRGs by using our CO and \md\ measurements.

A great source of uncertainty in this investigation is the derivation of \Mmol\ estimates that are dependent on the adopted CO-to-\mgas\ conversion factor ($\alpha_{\rm CO}$), which is known to vary by a factor of 6 between local normal galaxies and local ULIRGs (e.g., Downes \& Solomon 1998, Bolatto et al.\,2013). Here, instead of using a single $\alpha_{\rm CO}$ value for all sources, we take advantage of the dust mass estimates for our galaxies and employ the dust to gas mass ratio technique (e.g., Magdis et al.\,2012b). Adopting a solar metallicity,  and the dust to gas mass ratio versus metallicity relation ($\delta_{\rm GDR}-Z$) of Leroy et al.\,(2011), we find an \mgas/\md\ ratio of $\sim$80. Then using the \md\ estimates for our sample we derive the gas mass and the corresponding error for our sources, taking into account the \md\ uncertainties, as well as assuming a 0.2\,$dex$ uncertainty in the adopted metallicity and therefore in the assumed \mgas/\md\ ratio. A full description of the limitations inherent in this approach are presented in Magdis et al.\,(2012b). 

In Fig. \ref{fig:gas}\,(left) we plot the \lcii/\lfir\ ratio as a function of \lir/\Mmol {\footnote{We assume $M_{\rm HI} <<$ \Mmol, so that \mgas\ = \Mmol\ + $M_{\rm HI} \approx$ \Mmol}, focusing on the sample for which we have [CII] and [CO] 
measurements. This figure reveals a clear trend, with sources that deviate more than 1\,$\sigma$ from the \lcii-\lfir\ relation of eq.\,(3) exhibiting the highest star formation efficiencies. A Kendal's tau ranking test yields a correlation coefficient of $\rho = -0.9$ and a $p$-value of 4$\times$10$^{-5}$, suggesting a statistically significant anti correlation between the two parameters. For comparison we also include a sample of local disks and local ULIRGs adopting an $\alpha_{\rm CO}$ \footnote{The units of $\alpha_{\rm CO}$, M$_{\odot}$ pc$^{-2}$ (K km s$^{-1}$)$^{-1}$ are omitted from the text for brevity} conversation factor of 4.5 and 0.8, respectively (e.g., Downes \& Solomon\,1998, Rosolowsky et al.\,2007, Bolatto et al.\,2008,\,2013 Sliwa et al.\,2012) The majority of intermediate redshift (U)LIRGs in our sample fall in the locus of local disks, while the two [CII]-deficient sources have SFEs similar to that of local ULIRGs. Similar to Gracia-Carpio et al.\,(2011), we conclude that low \lcii/\lfir\ values are directly linked to high star formation efficiencies.

Using the derived \Mmol\ values from the $\delta_{\rm GDR}-Z$ method and the CO measurements we can also estimate the $\alpha_{\rm CO}$ values using the CO-to-$M_{\rm H_{\rm 2}}$ formula: 
\begin{equation}
\begin{centering}
\rm M_{\rm H_{\rm 2}} = \alpha_{\rm CO} \times L^{\prime}_{\rm CO}
\end{centering}
\end{equation}

\noindent The inferred $\alpha_{\rm CO}$ values for the whole sample range from 6 to 1. [CII]-deficient sources  have a mean  $\langle \alpha_{\rm CO} \rangle$ = 1.5, comparable to that of local ULIRGs. On the other hand, sources that  follow the \lcii-\lfir\ relation of eq.\,3 have a mean $\langle \alpha_{\rm CO} \rangle$ = 5.3, similar to that of local normal galaxies. 

Given the uncertainties coupled with the \Mmol\ estimates, to ensure that the result is not an artefact of the derived $\alpha_{\rm CO}$ values 
in Fig. \ref{fig:gas}\,(right), we plot  the \lcii/\lfir\ ratio as a function of the direct observables \lir/\lco. In this plot we also add a sample of high$-z$ ULIRGs, along with the samples presented in the \lcii/\lfir$-$\lir/\Mmol\ plot. Again, we find that there is a strong anti-correlation between the two ratios, with an increasing \lir/\lco\ for decreasing \lcii/\lfir, at all redshifts and luminosities. 

\subsection{The physical origin of CII emission}
It is worth to attemptimg to put constraints on 
its spatial origin, namely whether it predominantly originates from PDRs or from ionised HII-regions. For this task, ionized nitrogen [NII] 205$\,\mu$m can provide 
crucial information. With an  ionization potential  of 14.53 eV,  slightly above that of hydrogen, [NII] emission traces HII regions. On the other hand, the 
ionization potential of carbon is slightly below that of hydrogen, and consequently can arise both from HII regions and from PDRs. As [NII] and [CII] have very 
similar critical densities for excitation in ionized gas regions, their line ratio is insensitive to the hardness of the radiation field. Thus, for a given metallicity, their flux ratio can be used as a diagnostic tool to estimate the fraction of carbon emission arising from the HII regions (where both [CII] and [NII] are present) versus that 
arising from the neutral phase (where  [NII] is absent). In a HII region with fixed metallicity, the expected ratio of the two lines is practically independent of the 
gas density and  assuming  gas phase abundances of $n$([NII])/$n_{\rm e}$ = 7.9 $\times 10^{-5}$ and $n$([CII])/$n_{\rm e}$ = 1.4 $\times 10^{-4}$ (Savage \& Sembach\,1996), we expect a ratio of line luminosities  \lcii/
$L_{\rm [NII]}$ $\approx$ 2$-$3 in the case that both lines originate from ionised gas regions (Oberst et al.\,2006, Decarli et al.\,2014).

Since we do not have [NII] detection for any of our sources, we can only place lower limits in the  \lcii/$L_{\rm [NII]}$ ratio of our sample by using the [CII] 
measurements and the [NII] upper limits.  
While based on the lower limits we cannot decisively quantify the fraction of [CII] emission arising from the PDRs,  for the majority of the sources the 
\lcii/$L_{\rm [NII]}$ lower limits are well above the theoretical expectations (\lcii/$L_{\rm [NII]}$ $>$ 3.2) for  the case where [CII] emission originates purely from HII regions. In particular, it 
appears that  $>$50\%  of [CII] emission in intermediate redshift ULIRGs arises from PDRs, similar to high$-z$ ULIRGs (Decarli et al.\,2014) and local starbursts (Rigopoulou et al.\,2013).

\section{Discussion}
So far we have seen that the far-IR properties of intermediate redshift ULIRGs (0.2 $< z <$ 0.8) in our sample appear to deviate from those of local ULIRGs, while their overall characteristics resemble those of local normal galaxies and high$-z$ star forming ULIRGs. In particular, we showed that 
they exhibit \lcii/\lfir\ ratios very similar to that of local spirals and high$-z$ star forming galaxies, with almost an order of magnitude more 
luminous [CII] emission compared to the [CII]-deficient local (U)LIRGs. This observation points towards a universal \lcii$-$\lfir\ 
relation, with a constant \lcii/\lfir\ ratio at all epochs and luminosities (albeit with a considerable scatter), from which at least local ULIRGs 
and high$-z$ AGN dominated sources are strong outliers. We have also shown that at all luminosities and redshifts, sources follow a tight relation 
between their far-IR color and their fractional [CII] emission (\lcii/\lfir). The colder dust temperatures of high$-z$ and intermediate redshift ULIRGs, as 
found in the current study, offer an explanation for the observed discrepancy between the strength of [CII] emission of local and 
distant ULIRGs. Furthermore, we have shown that, as already reported for local galaxies by Gracia-Carpio et al.\,(2011), among intermediate 
redshift ULIRGs, sources with higher star-formation efficiencies tend to exhibit lower \lcii/\lfir\ ratios. Having compared the \lcii/\lfir\ ratios and dust temperatures in ULIRGs and normal galaxies across cosmic time (using our sample as well as literature data), the next step is to compare their star-formation efficiencies. Rather than using SFR and gas mass surface densities, we consider the simpler relations between integrated quantities, namely the total SFR and molecular gas mass or the corresponding observables, \lir\ and \lco.
\begin{figure*}
\centering
\includegraphics[scale=0.34]{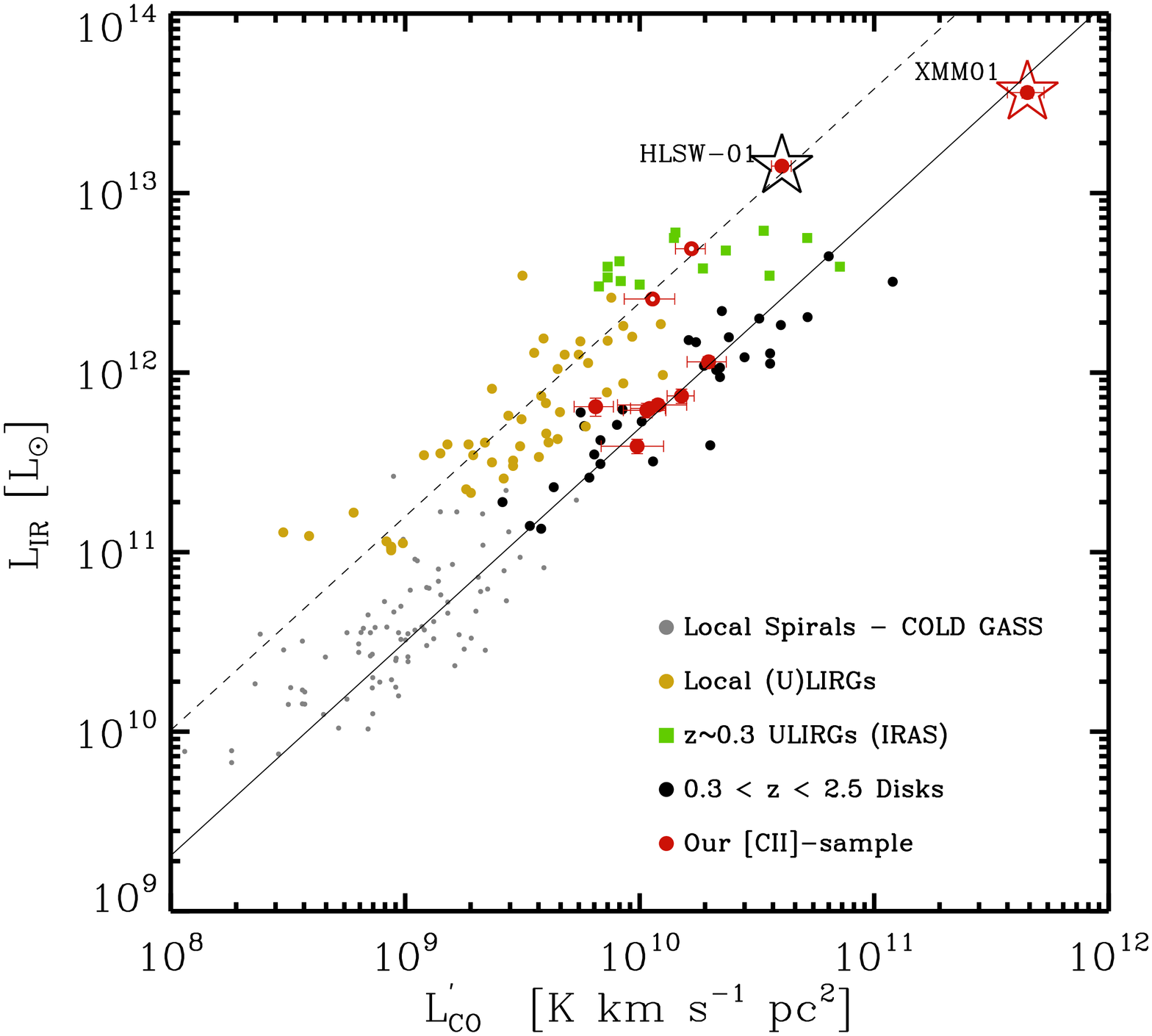}
\includegraphics[scale=0.34]{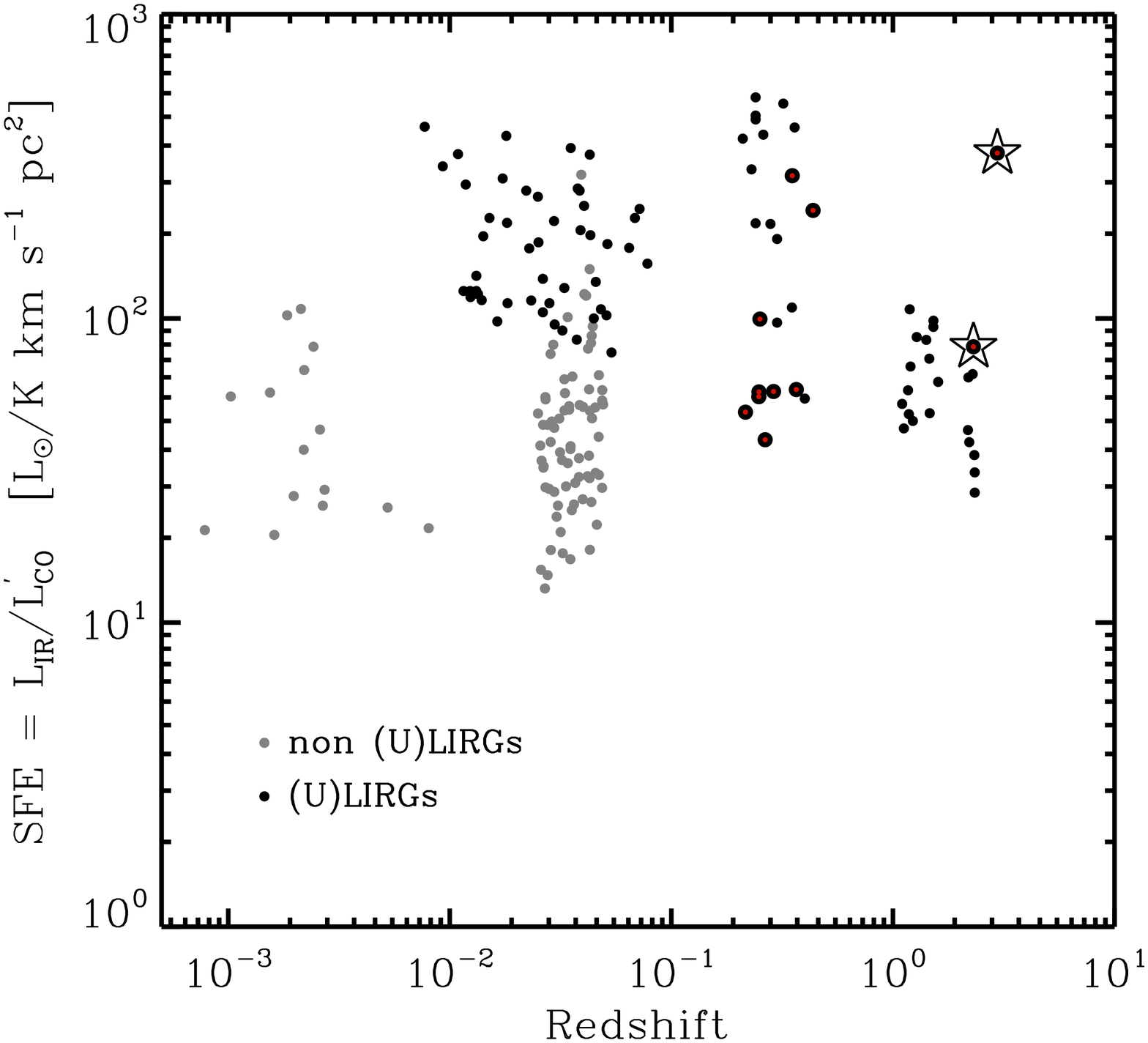}
\caption{{\textbf {Left:}} \lir\ versus \lco\ for galaxies in our sample that have CO and [CII] observations (filled red squares). [CII]-deficient sources are indicated with a small white circle. Gray circles are local normal galaxies from the COLD GASS survey, orange circles are local (U)LIRGs, green squares are $0.2 < z < 0.6$ and black circles $0.3 < z < 2.5$ star forming galaxies. The solid and dashed line represent the \lir-\lco\ disk and starbursts sequence of Sargent et al.\,(2013).  {\textbf {Right:}} Star-formation efficiency \lir/\lco\ as a function of redshift for the same samples as in the left panel, split between (U)LIRGs (\lir\ $>$ $10^{11}$\,\lsol, black circle)and non-ULIRGs (\lir\ $<$ $10^{11}$\,\lsol, grey circles). Our sample of intermediate redshift (U)LIRGs are indicated with a filled red circle.}
\label{fig:sfe} %
\end{figure*}

In Fig.  \ref{fig:gas}(left) we compare the CO and IR luminosities of our [CII]-sample  with other cosmologically relevant galaxy populations:  local spirals (COLD GASS, Saintonge et al.\,2011); local ULIRGs (Solomon et al.\,1997); $z \sim 0.3$ disks (Geach et al.\,2011) {\textit{IRAS}} selected $z 
\sim 0.3$ ULIRGs (Combes et al 2011); and high$-z$ star forming galaxies (Daddi et al.\,2010, Genzel et al.\,2010).  The majority of $z 
\sim0.3$ (U)LIRGs in our sample follow the  \lco$-$\lir\ relation of 
normal galaxies from Sargent et al.\,(2013) which is offset by 0.46\,$dex$ with respect to that of local and intermediate redshift {\textit{IRAS}} selected ULIRGs, suggesting that in $z \sim 0.3$ \h-selected ULIRGs CO emission is  $\sim$3 more luminous for a given \lir. Interestingly, the two sources in our sample that deviate the most from the sequence of disks are those with the lowest \lcii/\lfir\ ratio. Similarly, in  Fig. \ref{fig:gas}\,(right) we plot the star-formation efficiency SFE = \lir/\lco\ (in units of \lsol /[K km s$^{-1}$ pc$^{2}$]) as a function of redshift for the same samples as in the left panel. The ULIRGs in our sample exhibit star-formation efficiencies comparable to that of high$-z$ star-forming galaxies and local spirals, with  two exceptions  being the strong outliers of the \lcii$-$\lfir relation. In particular, for the whole  sample we find a median star formation efficiency of 58  \lsol /[K km s$^{-1}$ pc$^{2}$] compared to $\sim$ 300 for {\textit{IRAS}} ULIRGs, $\sim$ 200 for local ULIRGs, $\sim$ 40 for local spirals and $\sim$ 50 for high$-z$ ULIRG-like disks. 

Evidently, \h\ has revealed a population of intermediate redshift ULIRGs with large molecular gas reservoirs that can maintain their high star formation rates for 
prolonged time periods. Indeed, using a typical $\alpha_{\rm CO}$ value of 4.5, as derived from the previous analysis, as well as the Kennicutt (1998) formula to 
convert the \lir\ to SFR (SFR $=$ 1 $\times$ 10$^{-10}$ \lir, for a Chabrier IMF), we find a gas depletion time scale ($\tau_{\rm dep}$ = \Mmol/SFR) of $\sim$ 0.8 Gyrs, a factor of 10 longer than for  the short-lived starburst events of local ULIRGs ($\sim$ 80\,Myrs). We thus find that the evolution of the star formation activity  between a long lasting star formation in a large fraction of high$-z$ ($ z > 1$) ULIRG-like star forming galaxies and a short-lived starbursts event in the majority of the local ULIRGs reported in various studies (Genzel et al.\,2010, Tacconi et al.\,2010, Daddi et al.\,2010, Magdis et al.\,2012) appears to manifest itself already at moderate cosmological times (i.e., 3$-$4 billion years ago). 


An explanation for the observed evolution offered by recent theoretical studies is a shift from a merger induced to  more quiescent star formation activity (e.g., Dekel et al.\,2009, Dav\'e et al.\,2010). Recent morphological and kinematical studies have revealed an increasing number of non-interacting, disk dominated ULIRG-like systems with look back time (e.g., Forster 
Schreiber et al.\,2009, Kartaltepe et al.\,2011, Wang et al.\,2012). Also, as discussed above, high$-z$ star-forming galaxies exhibit colder SEDs (e.g., Swinbank et al.\,2010, Hwang et al.\,2010, Magdis et al.\,2010, Symeonidis et al.\,2013)  compared to local ULIRGs and more extended star formation (e.g.,  Men\'endez-Delmestre et al.\,2009, Rujopakarn et al.\,2011). Finally, galaxies at all redshifts appear to follow a tight correlation between their star-formation rate and they stellar mass (e.g., Noeske et al.\,2007, Elbaz et al.\,2007, Panella et al.\,2009, Daddi et al.\,2010, Magdis et al.\,2010, Karim et al.\,2011). This speaks against stochastic merger induced 
star forming episodes as the main driver of star formation activity, as they would introduce a large scatter in the SFR$-$$M_{\ast}$ relation.

Rigopoulou et al.\,(2014) using the current sample of intermediate redshift ULIRGs modelled their \lcii/\lfir\ and \lir/\lco\ ratios and provided evidence that their star forming regions are illuminated by moderately intense FUV radiation, with $G_{0}$ in the range 10$^{2}$$-$10$^{2.5}$ and ${\it G_{0}}/ \eta \sim$0.1$-$1 cm$^{3}$  similar to that of local normal galaxies (e.g. Wolfire et al.\,1990) and high$-z$ star forming galaxies (Stacey et al.\, 2010). On the other hand, local ULIRGs exhibit stronger radiation fields with $G_{0} \sim 10^{2.2-3.6}$ (Farrah et al.\,2013). Also,  Diaz-Santos et al.\,(2013),\,(2014) recently revealed a clear trend  between \lcii/\lfir\ and the compactness of the mid-IR emission of the local galaxies; for a given \lir, sources with lower \lcii/\lfir\ ratios are warmer and their  hot dust is confined towards a smaller volume in the centre of the galaxy, suggesting more compact star forming regions. They also found that [CII]-deficient regions in local LIRGs are restricted to the nuclei of the source. The high \lcii/\lfir\  ratios of our sample and the  colder dust temperature, both indicative of extended star-formation activity together with the large molecular gas reservoirs and the long gas depletion time scales, signify a strong evolution in ULIRGs even at moderate redshifts, suggesting  that the nature of the ULIRG population changes very fast from exclusively compact merger-driven to a more varied population.

The observed steep increase of the molecular gas fraction in star-forming galaxies with look-back time (e.g. Geach et al.\,2011, Magdis et al.\,2012a,b) can explain both their high star formation rates and their long gas depletion time scales, without the need of a merger-induced star forming episode. Indeed, based on the evolution of gas fraction from Magdis et al.\,(2012), galaxies by $z \sim0.5 $ already have  two times more molecular gas compared to those in the local Universe. While all the above provide substantial evidence, a  kinematical study of the current sample using IFU spectroscopy with SWIFT (Houghton et al.\,in prep) as well high resolution HST imaging that will trace the morphology of the current sample, are best suited to address whether the observed evolution is indeed driven by a change in the triggering mechanism of star formation or by other processes, such an evolution in the galactic scale metal content of ULIRGs with cosmic time.

\section{Conclusions}
Using \h-FTS far-IR spectroscopy, CO observations and detailed mid-to-far-IR photometry from \s\ and \h\ we have preformed a detailed investigation of the properties of 17 \h-selected intermediate redshift ($0.2 < z < 0.8$) (U)LIRGs. The current sample complements the \td-biased and luminosity limited {\textit{IRAS}} sample of ULIRGs at this redshift range. By combining unique [CII] measurements, with molecular gas observations and detailed spectra energy distributions we reach the following conclusions.  

\begin{enumerate}
\item The majority of \h-selected intermediate redshift ULIRGs have luminous [CII] emission lines exhibiting \lcii/\lfir\ ratios that are a factor of 10 higher than that of local ULIRGs and comparable to that of local normal and high-$z$ star forming galaxies. By combining our sample with data from the literature we find that the majority of galaxies at all redshift, and all luminosities follow a \lcii$-$\lfir\ relation with a slope of unity and a scatter of 0.3\,$dex$. At least local ULIRGs and high$-z$ AGN dominated sources are strong outliers from this relation.

\item We find a strong anti correlation between the \lcii/\lfir\ ratio and the far-IR color $L_{\rm 60}$/$L_{\rm 100}$, in the sense that warmer sources exhibit lower  \lcii/\lfir, at all redshifts and all luminosities.  Intermediate redshift (U)LIRGs similar to high$-z$ ULIRG like star forming galaxies exhibit colder dust temperatures.

\item Our sample exhibits lower star formation efficiencies compared to local ULIRGs and large molecular gas reservoirs. Sources with higher star formation efficiencies tend to have lower \lcii/\lfir\ ratios.

\item The high \lcii/\lfir\ ratios, the moderate star formation efficiencies (\lir/\lco\ or \lir/\Mmol) and the relatively low dust temperatures of our sample indicate that the observed evolution of the ULIRG phenomenon between $z = 0$ and $ z > 1$ is already taking place by $z \sim 0.3$.  

\end{enumerate}

\textit{Acknowledgments.} We warmly thank the referee for his
/her constructive comments and suggestions. 
GEM and DR acknowledge support from STFC through grant ST/K00106X/1. GEM acknowledges support from the John Fell Oxford University Press (OUP) Research Fund and the University of Oxford. A.A.-H. acknowledges funding through the Universidad de Cantabria
August G. Linares Programme. This work includes observations made with IRAM, which is supported by INSU/CNRS (France), MPG (Germany), and IGN (Spain).  Based on observations made with ESO Telescopes at the APEX Observatory under programme ID  090.B-0708A and  091.B-0312A.  
This paper uses data from \h's photometer SPIRE. SPIRE has been developed by a consortium of institutes led by Cardiff University (UK) and including: Univ. Lethbridge (Canada); NAOC (China); CEA, LAM (France); IFSI, Univ. Padua (Italy); IAC (Spain); Stockholm Observatory (Sweden); Imperial College Lon- don, RAL, UCL-MSSL, UKATC, Univ. Sussex (UK); and Caltech, JPL, NHSC, Univ. Colorado (USA).

\clearpage

\end{document}